\documentclass[12pt]{iopart}

\usepackage{graphicx}

\newcommand{\bra}[1]{\langle#1|}
\newcommand{\ket}[1]{|#1\rangle}
\newcommand{\bracket}[2]{\big\langle#1 \bigm| #2\big\rangle}

\renewcommand{\Tr}{{\rm Tr}}
\renewcommand{\Im}{{\rm Im}}

\newcommand{\p}{{\prime}}

\newcommand{\Mat}[1]{{\bf #1}}
\newcommand{\Span}{{\rm span}}

\begin{document}

\title[Ab initio theory of strong correlations in nanoscale devices]{
  Towards a full ab initio theory of strong electronic correlations in nanoscale devices
}

\author{David Jacob}
\address{Max-Planck-Institut f\"ur Mikrostrukturphysik, Weinberg 2, 06120 Halle, Germany}
\ead{djacob@mpi-halle.de}

\begin{abstract}
  In this paper I give a detailed account of an ab initio methodology for describing 
  strong electronic correlations in nanoscale devices hosting transition metal atoms
  with open $d$- or $f$-shells. The method combines Kohn-Sham Density Functional Theory 
  for treating the weakly interacting electrons on a static mean-field level with 
  non-perturbative many-body methods for the strongly interacting electrons in the open 
  $d$- and $f$-shells.
  An effective description of the strongly interacting electrons in terms of a multi-orbital 
  Anderson impurity model is obtained by projection onto the strongly correlated subspace 
  properly taking into account the non-orthogonality of the atomic basis set. 
  A special focus lies on the ab initio calculation of the effective screened interaction matrix 
  U for the Anderson model.
  Solution of the effective Anderson model with the One-Crossing approximation
  or other impurity solver techniques yields the dynamic correlations within the 
  strongly correlated subspace giving rise e.g. to the Kondo effect.
  As an example the method is applied to the case of a Co adatom on the Cu(001) surface.
  The calculated low-bias tunnel spectra show Fano-Kondo lineshapes 
  similar to those measured in experiments. 
  The exact shape of the Fano-Kondo feature as well as its width 
  depend quite strongly on the filling of the Co $3d$-shell.
  Although this somewhat hampers accurate quantitative predictions regarding lineshapes 
  and Kondo temperatures, the overall physical situation can be predicted quite reliably.
\end{abstract}

\section{Introduction}

Modern experimental techniques now allow to reliably create, manipulate and control
nanoscale devices with atomic precision in the lab thus bringing the dream of
molecular electronics or nanoelectronics to create ultimately miniaturized electronic devices from 
single molecules closer to reality~\cite{Heath:PT:2003,Joachim:PNAS:2005,Cuniberti::2005,Cuevas::2010}.
Prospective building blocks for molecular electronic circuits such as molecular 
rectifiers~\cite{Aviram:CPL:1974,Elbing:PNAS:2005} and field-effect transistors~\cite{Tans:N:1998,Martel:APL:1998} 
have already been demonstrated in experiments.
The use of magnetic atoms or molecules promises to further enhance the functionality of molecular
devices by exploiting the spin-degree of freedom of the electron in addition to its charge. Such
devices could serve e.g. as basic building blocks for nanoscale spintronics 
applications~\cite{Zutic:RMP:2004,Bogani:NM:2008} or as ultimately miniaturized magnetic information 
storage devices~\cite{Gatteschi::2006}.

Naturally, quantum effects play a crucial role in electronic devices of such tiny dimensions. 
Consequently, experiments with atomic- and molecular-scale devices have produced a wealth of 
quantum phenomena such as conductance quantization~\cite{Agrait:PR:2003}, 
quantum interference~\cite{Liang:N:2001,Kong:PRL:2001}, or quantum phase transitions~\cite{Roch:N:2008}.
On the other hand, details of the atomic structure also play an important role for determining 
the electronic properties of nanoscale devices, especially regarding the contact between
molecule and metal leads~\cite{Palacios:PRL:2003,Palacios:PRB:2008,Schull:PRL:2009}.
Also the coupling to the leads can significantly alter the electronic and magnetic 
properties of nanoscale devices by broadening and shifting of energy levels, as well as screening
effects. Hence a proper theoretical description of nanoelectronic devices needs to take
into account all of the following: quantum effects, the actual atomic structure of the 
device and the coupling to the leads. 


The now standard approach for the description of molecular electronic devices is to combine
density functional theory (DFT) calculations with the Landauer transport theory or with the 
non-equilibrium Green's function formalism (NEGF)~\cite{Taylor:PRB:2001a,Palacios:PRB:2002,Brandbyge:PRB:2002,Rocha:PRB:2006}. 
The DFT based transport approach yields an effective mean-field description for the electronic 
structure and transport properties of molecular devices, taking into account quantum effects, 
as well as the actual atomic structure of the device, and the coupling of the device to the 
metallic leads. The approach works quite well for the description of metallic nanocontacts and 
nanowires and carbon nanotubes~\cite{Brandbyge:PRB:2002,Mehrez:PRB:2002,Palacios:PRL:2003}.
On the other hand, it was realized quite early on that this approach often overestimates conductances
of molecules attached to metal leads by orders of magnitude. Its origin has been a matter of debate 
for over a decade and is still not completely settled~\cite{DiVentra:PRL:2000,Varga:PRL:2007,Lindsay:AM:2007}. 


Moreover, nanoscale devices comprising magnetic atoms or molecules often display phenomena 
induced by so-called strong dynamic correlations that arise when the effective Coulomb 
interaction between the electrons exceeds their kinetic energies. Dynamic correlations 
can have a profound impact on the electronic and magnetic structure and the transport properties 
of the system. One of the most intriguing phenomena induced by dynamic correlations in 
nanoscale devices is probably the Kondo effect~\cite{Kondo:PTP:1964,Hewson::1997}: 
Below a critical temperature characteristic of the system, the Kondo temperature $T_K$, 
the atomic or molecular spin forms a many-body singlet state with the nearby conduction electrons,
thereby screening the magnetic moment of the device. 
The correlations usually originate from the strongly interacting open $3d$- or $4f$-shells of 
transition metal atoms. But also molecular orbitals of purely organic molecules only weakly 
coupled to the leads can give rise to strong correlations. This is corroborated by the fact 
that the Kondo effect is not only frequently observed in molecular devices comprising transition 
metal atoms~\cite{Madhavan:S:1998,Li:PRL:1998,Park:N:2002,Liang:N:2002,Zhao:S:2005,Yu:PRL:2005,
Iancu:NL:2006,Fu:PRL:2007,Calvo:N:2009,Franke:S:2011,Minamitani:PRL:2012,Kuegel:NL:2014},
but also for devices made from purely organic molecules~\cite{Nygard:N:2000,Park:N:2002,Yu:NL:2004,
Jarillo-Herrero:N:2005,Parks:PRL:2007,Roch:PRL:2009}.
By construction the DFT based transport method being a static mean-field approach 
cannot capture the dynamic correlations that lead e.g. to the Kondo effect in nanoscale 
devices~\footnote{
  Recently it has been shown by Bergfield {\it et al.} that the exact exchange correlation
  functional yields the exact transmission at the Fermi level in the case of the 
  simple Anderson impurity model. However, even the exact Kohn-Sham spectrum does
  not yield a correct description of the spectral function and transmission outside
  the Fermi level. Hence the renormalization of the Kondo peak by the interactions
  cannot be captured by Kohn-Sham DFT based transport calculations~\cite{Bergfield:PRL:2012}.
}
This neglect of dynamic correlations could also be behind the afore mentioned overestimate of the conductances 
of molecular devices by the DFT based transport approach since dynamic correlations can lead to a strong 
renormalization of the quasi particles relevant for the transport through the molecule~\cite{Mera:PRL:2010}.


Recent efforts to go beyond the DFT based transport approach are to combine time-dependent DFT 
(TDDFT) with the NEGF~\cite{Stefanucci:PRB:2004,DiVentra:JoPCM:2004} or the GW approximation with 
NEGF~\cite{Darancet:PRB:2007,Thygesen:JCP:2007,Darancet:NL:2012}. A problem of the TDDFT approach 
is that the standard approximations for TDDFT functionals in connection with an adiabatic 
exchange correlation kernel does not yield an improvement for the description of correlation effects
with respect to the static DFT approach. Some progress has been made recently in that direction
by finding a non-adiabatic exchange-correlation kernel for strongly correlated systems but only in the context 
of simplified models such as the Hubbard or Anderson model~\cite{Stefanucci:PRL:2011,Turkowski:JoPCM:2014}. 
The GW based transport approach on the other hand has been implemented in a fully ab initio way and has 
been applied to realistic molecular devices. Although GW yields an energy-dependent self-energy for describing
the electronic interactions and thus captures dynamic correlation effects to some extend, it is 
perturbative in nature and thus strong electronic correlations such as those leading to the Kondo effect 
or the Mott-Hubbard metal-insulator transition are not properly described. 

Here I give a detailed account of a different ab initio approach for the description of strongly 
correlated molecular conductors which has been developed, successively refined and extended in 
previous work~\cite{Jacob:PRL:2009,Jacob:PRB:2010,Jacob:PRB:2010a,Karolak:PRL:2011,Jacob:PRB:2013}. 
In this approach only the strongly interacting part of the electronic 
spectrum is described by advanced many-body methods in order to capture dynamic correlations effects. 
The weakly to moderately interacting part of the electronic system is still treated on a static 
mean-field level by standard Kohn-Sham DFT (KSDFT). This approach is basically an adaption of the 
DFT+Dynamical Mean-Field Theory (DFT+DMFT) 
approach~\cite{Anisimov:JPCM:1997a,Kotliar:RMP:2006,Karolak:JoPCM:2011,Pourovskii:PRB:2007}, which has been developed
for the realistic description of strongly correlated solids, to the special situation of nanoscale
conductors. Similar approaches for treating strong correlations in molecular devices have recently
appeared in the literature~\cite{DiasdaSilva:PRB:2009,Korytar:JoPCM:2011,Ishida:PRB:2012,Valli:PRB:2012,Ryndyk:PRB:2013,Baruselli:PRB:2013}.

This paper is organized as follows: In Sec.~\ref{sec:method} a detailed account of the 
so far developed methodology is presented. In Sec.~\ref{sec:results} the methodology is applied to the case
of a Co adatom at the Cu(001) surface which has been studied extensively in the recent 
past~\cite{Knorr:PRL:2002,Wahl:PRL:2005,Neel:PRL:2007,Uchihashi:PRB:2008} 
and thus presents an ideal testbed for the theory. In Sec.~\ref{sec:conclusions}, I draw conclusions from comparison of the 
results to the experiments and other theoretical methods. I also discuss some of the caveats of the developed 
theory and possible solutions to these problems as well as future directions.

\section{Methodology}
\label{sec:method}

\begin{figure}
  \begin{tabular}{ccc}
    (a) & (b) & (c) \\
    & & \\
    \includegraphics[width=0.35\linewidth]{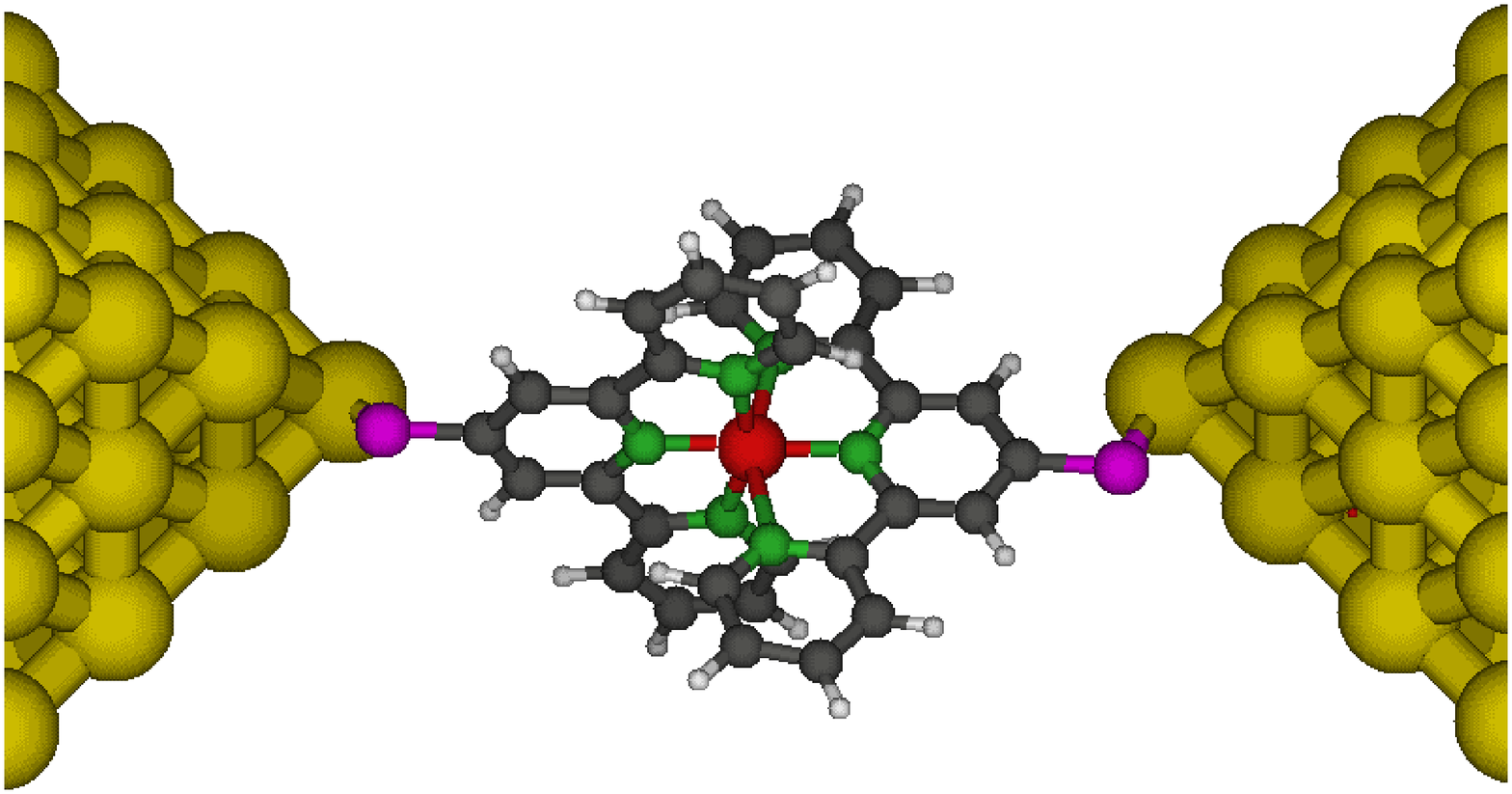} &
    \includegraphics[width=0.25\linewidth]{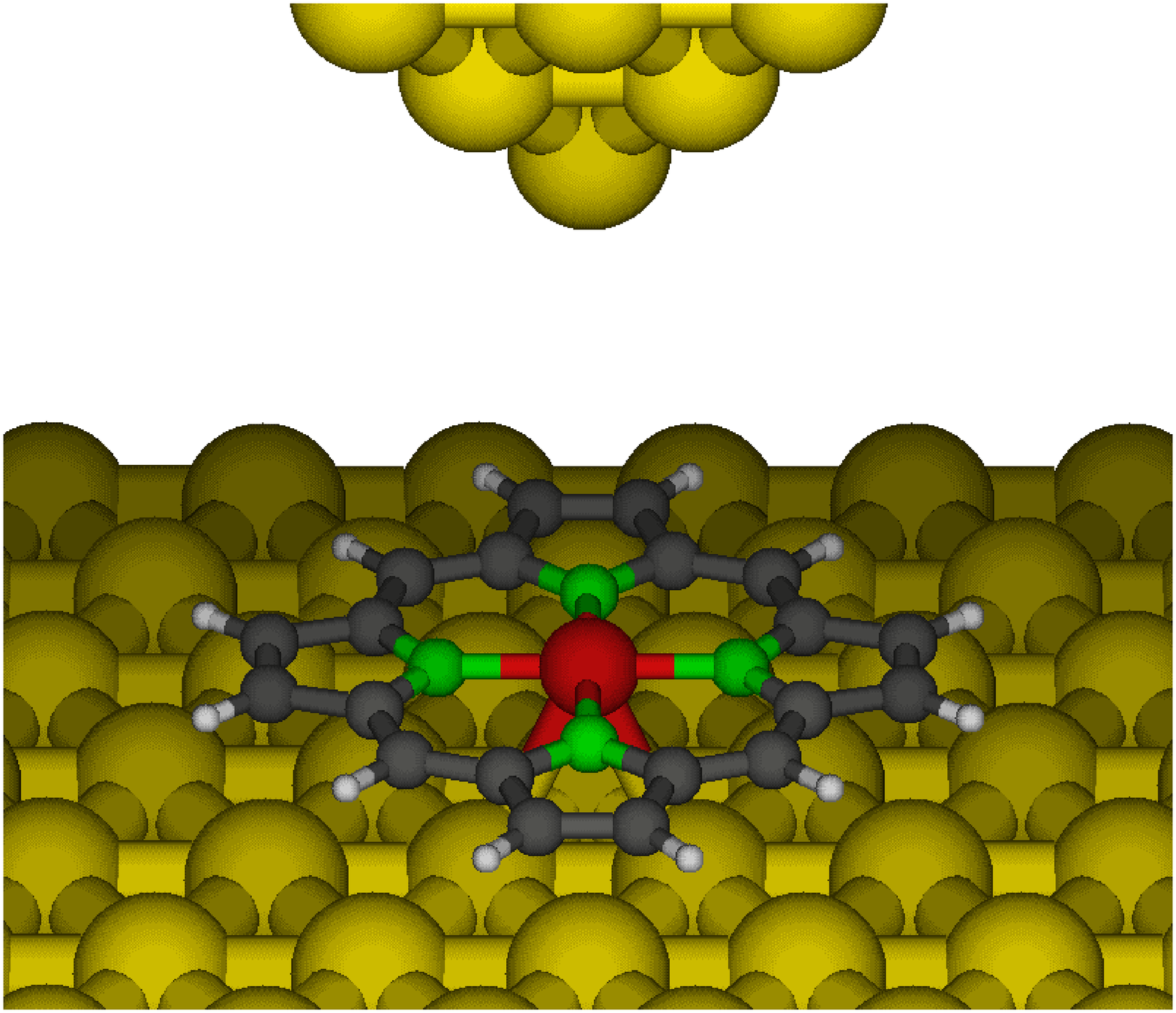} &
    \includegraphics[width=0.33\linewidth]{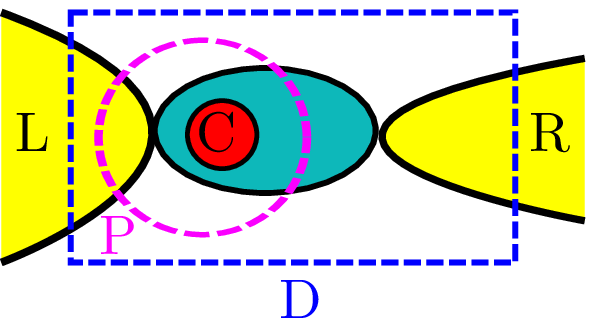}
  \end{tabular}
  \caption{
    \label{fig:scheme}
    Typical situations encountered in molecular electronics/spintronics:
    (a) A magnetic molecule bridging the tips of a nanocontact.
    (b) A magnetic molecule on a metal surface probed by an STM tip.
    (c) Schematic sketch of model that captures both situations shown in (a) and (b).
    A central atom or molecule (turquoise) hosting strongly correlated levels C (red)
    is connected to two metal leads L and R (yellow). The device region D (blue) 
    contains the central atom/molecule and part of the leads. The polarization 
    region P (magenta) extends over that part of the atom/molecule and the lead(s) 
    in close proximity to C.
  }
\end{figure}

The typical situations encountered in experiments with atomic and molecular devices are depicted in 
Fig.~\ref{fig:scheme}: (a) A magnetic molecule suspended between the tips of a metal nanocontact 
and (b) a magnetic atom or molecule deposited on a metallic surface probed by an STM tip. The magnetism 
and hence the strong correlations of the molecule are here assumed to stem from a single transition metal 
atom at its center. But the approach can be easily generalized to the case of multiple magnetic atoms by 
adaption of the Dynamical Mean-Field Theory (DMFT) to the case of molecular conductors~\cite{Jacob:PRB:2010a}. 
Both situations depicted in Figs.~\ref{fig:scheme}(a,b) can be described by the model depicted schematically 
in Fig.~\ref{fig:scheme}(c): the central region, called device region D, contains the molecule or atom and part 
of the leads, and will be described on the level of KSDFT.
Within the atom or molecule the correlated subspace C yields the strongly interacting levels of 
the atom/molecule that will be treated by advanced many-body techniques in order to capture the strong dynamic 
correlations. The polarization region P where the polarizability is calculated in order to compute the screened 
interaction $U$ of the strongly correlated subspace on the other hand extends over that part of the molecule and/or 
leads in immediate vicinity of the correlated subspace C. 
This approach can in principle also be applied directly to the case of purely organic molecules. In that case
one has to identify the molecular orbitals responsible for the strong correlations~\cite{Soriano::2014}.

The approach has been implemented within the ANT.G package~\cite{ANTG}
which interfaces the Gaussian quantum chemistry code~\cite{G09} in 
order to implement the DFT based ab initio transport methodology for
molecular conductors. The Gaussian code makes use of Gaussian atomic orbitals
as basis sets for performing quantum chemistry and DFT calculations
of finite clusters and molecules. The ANT.G package embeds the finite
cluster representing the device region into bulk electrodes in order
to model the transport situation depicted in Fig.~\ref{fig:scheme}.
However, the formalism developed below is not specific to Gaussian
basis sets. It can directly be applied to any atomic basis set, as for 
example the Fireball orbitals used in the SIESTA code~\cite{SIESTA}. 
Even more general, the formalism might be applied to any basis set
as long as the different subspaces (D,P and C) can be defined in a 
meaningful way.

\subsection{Non-orthogonal basis sets and projection onto a subspace}
\label{sub:nobs_projections}

We now have to carefully define the projections onto the different subspaces taking
into account the non-orthogonality of the atomic basis set. The choice of projection 
strongly influences physical quantities associated with the subspace such as the density and 
electronic occupancy of the subspace as has been shown recently by Soriano and Palacios
\cite{Soriano:PRB:2014}.

We assume that the Hilbert space ${\rm H}$ of our system is spanned by a (finite) 
set of non-orthogonal orbitals $H={\{\ket{\alpha}\}}$, i.e. ${\rm H}=\Span(H)$,
and $\bracket{\alpha}{\beta}=S_{\alpha\beta}\ne0$ for $\ket{\alpha},\ket{\beta}\in{}H$.
We now want to project onto a subspace M of H spanned by a subset $M=\{\ket{m}\}$
of the orbitals $\ket{\alpha}\in{H}$, i.e. $M\subset{H}$.
Due to the non-orthogonality of the orbitals
$\ket{\alpha}\in{H}$, subspace M will in general have a finite overlap with the 
subspace R spanned by the rest of the orbitals $\ket{r}\in{R}\equiv{}H\setminus{M}$, i.e.
$S_{mr}=\bracket{m}{r}\ne0$ for $\ket{m}\in{M}$ and $\ket{r}\in{R}$.
Hence the question arises how to define 
a proper projection $\hat{P}_{\rm M}$ onto that subspace. 
We note that there has actually 
been some controversy about this question in the literature 
(see e.g. Ref. \cite{ORegan:prb:2011} and references therein). 

It turns out that the proper choice for $\hat{P}_{\rm M}$ is actually quite obvious:
Let us first consider the simplest case of the subspace M being spanned by 
a single orbital $\ket{m}$. By definition, the projection operator for a single state 
is simply $\hat{P}_m=\ket{m}\bra{m}$. This definition is \emph{independent} 
of how (in which basis) the Hilbert space of the entire system is defined; i.e. it does not matter 
whether $\ket{m}$ forms part of the basis set spanning the entire Hilbert space or not;
or in case it does whether it has some overlap with the Hilbert space R spanned
by the rest of the basis set. 

Hence it is clear that the projection $\hat{P}_{\rm M}$ for the subspace M
can be written in an \emph{orthonormal} basis set $M^\perp=\{\ket{m^\perp}\}$ spanning the subspace 
M as $\hat{P}_{\rm M} = \sum_{m^\perp\in{M^\perp}} \ket{m^\perp}\bra{m^\perp}$. Such an orthonormal
set can always be found by L\"owdin orthogonalization of the original non-orthogonal 
set spanning M: $\ket{m^\perp}=\sum_m (\Mat{S}^{-1/2}_{\rm M})_{m{}m^\perp}\ket{m}$
where $\Mat{S}_{\rm M}$ is the overlap matrix between the basis set elements of M only
and $\Mat{S}^{-1/2}_{\rm M}$ is an abbreviation for $(\Mat{S}_{\rm M})^{-1/2}$, i.e. 
the matrix power $-1/2$ of the matrix $\Mat{S}_{\rm M}$. 
Hence we find for the projection operator:
\begin{eqnarray}
  \label{eq:Projector}
  \hat{P}_{\rm M} &=& \sum_{m^\perp\in{M^\perp}} \ket{m^\perp}\bra{m^\perp}
  = \sum_{m,n\in{M}} \sum_{m^\perp\in{M^\perp}} (\Mat{S}^{-1/2}_{\rm M})_{m{}m^\perp} (\Mat{S}^{-1/2}_{\rm M})_{m^\perp{n}} \ket{m} \bra{n}
  \nonumber\\
  &=& \sum_{m,n\in{M}} \ket{m} (\Mat{S}^{-1}_{\rm M})_{mn} \bra{n}
\end{eqnarray}
which is nothing but the identity operator for the subspace M written 
in the non-orthogonal basis set. 
It has been argued on more formal grounds that this choice for the projection is
actually the only physical reasonable one as it is the only one that leads to 
a tensorial consistent occupancy matrix which generates a Hermitian potential
\cite{ORegan:prb:2011}. 
Note that the subspace projection $\hat{P}_{\rm M}$ defined here corresponds to 
the projector with regard to the  $\Delta$ metric denoted by $\hat{P}_{\rm M}^\Delta$ 
in Ref.~\cite{Soriano:PRB:2014}.

Also note that in general we cannot write the identity operator for the entire system as the 
sum of the projection onto subspace M and subspace R spanned by the rest of the
basis set if there is some overlap between the two subspaces, i.e. 
$\hat{I}\neq\hat{P}_{\rm M}+\hat{P}_{\rm R}$. 
Rather we have to correct for the overlap between the two subspaces:
\begin{equation}
  \label{eq:Id}
  \hat{I} = \sum_{\alpha,\beta\in{H}} \ket{\alpha} (\Mat{S}^{-1})_{\alpha\beta} \bra{\beta} 
  = \hat{P}_{\rm M}+\hat{P}_{\rm R} + \hat{O}
\end{equation}
where $\Mat{S}^{-1}$ is the inverse of the overlap matrix for the \emph{entire} system.
$\hat{O}$ is an operator correcting the sum of projections by the overlap between 
the two subspaces M and R. $\hat{P}_{\rm \bar{M}}\equiv\hat{P}_{\rm R}+\hat{O}$ defines
the projection onto a new subspace $\bar{\rm M}$ which is actually orthogonal to subspace M.
The projection $\hat{P}_{\rm \bar{M}}$ thus defines an orthogonalization scheme which 
orthogonalizes R with respect to subspace M preserving the latter.

Now let us have a look at how an operator $\hat{A}$ acting on the full Hilbert space
is projected onto the subspace M:
\begin{eqnarray}
  \label{eq:OpProjection}
  \hat{A}_{\rm M} &\equiv& \hat{P}_{\rm M}\hat{A}\hat{P}_{\rm M}
  = \sum_{m,m^\p,n,n^\p\in{\rm M}} \ket{m} (\Mat{S}^{-1}_{\rm M})_{mm^\prime} 
  \bra{m^\prime} \hat{A} \ket{n^\prime} (\Mat{S}^{-1}_{\rm M})_{n^\prime{}n} \bra{n}
  \nonumber\\
  &=& \sum_{m,n\in{\rm M}} \ket{m} (\Mat{S}^{-1}_{\rm M}\Mat{A}_{\rm M}\Mat{S}^{-1}_{\rm M})_{mn} \bra{n}
  = \sum_{m,n\in{\rm M}} \ket{m} (\widetilde{\Mat{A}}_{\rm M})_{mn} \bra{n}
\end{eqnarray}
where $\Mat{A}_{\rm M}=(\bra{m}\hat{A}\ket{n})$ is the direct matrix given
by the matrix elements of $\hat{A}$ with the basis $\{\ket{\alpha}\}$
of subspace M, and $\widetilde{\Mat{A}}_{\rm M}=\Mat{S}^{-1}_{\rm M}\Mat{A}_{\rm M}\Mat{S}^{-1}_{\rm M}$ 
is the so-called nuclear matrix in that basis. Note that for an orthonormal basis 
of M we have $\widetilde{\Mat{A}}_{\rm M}=\Mat{\Mat{A}}_{\rm M}$.

Frequently, we will also have to project an operator $\hat{A}$ given for 
some subspace M onto a smaller subspace ${\rm M}^\prime\subset{\rm M}$:
\begin{eqnarray}
  \label{eq:OpSubProjection}
  \hat{A}_{\rm M^\prime} &=& \hat{P}_{\rm M^\prime} \hat{A}_{\rm M} \hat{P}_{\rm M^\prime}
  = \sum_{m,n\in{\rm M}} \hat{P}_{\rm M^\prime} \ket{m} (\widetilde{\Mat{A}}_{\rm M})_{\alpha\beta} \bra{n} \hat{P}_{\rm M^\prime} 
  \nonumber\\
  &=& \sum_{{m,n\in{\rm M}}\atop{m^\prime,n^\prime,p^\prime,q^\prime\in{\rm M^\prime}}}
  \ket{m^\prime} (\Mat{S}_{\rm M^\prime}^{-1})_{m^\prime{}p^\prime} \bracket{p^\prime}{m}
  (\widetilde{\Mat{A}}_{\rm M})_{mn} \bracket{n}{q^\prime} (\Mat{S}_{\rm M^\prime}^{-1})_{q^\prime{}n^\prime} \bra{n^\prime}
  \nonumber\\
  &=& \sum_{m^\prime,n^\prime\in{\rm M^\prime}}  \ket{m^\prime} (\Mat{S}_{\rm M^\prime}^{-1} \, \Mat{S}_{\rm M^\prime{}M} \, \widetilde{\Mat{A}}_{\rm M} \, 
  \Mat{S}_{\rm MM^\prime} \, \Mat{S}_{\rm M^\prime}^{-1})_{m^\prime{}n^\prime} \bra{n^\prime} 
\end{eqnarray}
where $\Mat{S}_{\rm M^\prime{}M}$ is the overlap matrix between orbitals $\ket{m^\prime}\in{M^\prime}$
and orbitals $\ket{m}\in{M}$
Hence we obtain the following expression for the nuclear matrix of subspace M${}^\prime$
in terms of the nuclear matrix for subspace M:
\begin{equation}
  \label{eq:MatSubProjection}
  \widetilde{\Mat{A}}_{\rm M^\prime} = \Mat{S}_{\rm M^\prime}^{-1} \, \Mat{S}_{\rm M^\prime{}M} \, \widetilde{\Mat{A}}_{\rm M} \,
  \Mat{S}_{\rm MM^\prime} \, \Mat{S}_{\rm M^\prime}^{-1}  
\end{equation}

On the other hand, we may also have the opposite situation where we have some operator $\hat{A}_{\rm M}$ 
only defined on subspace M, and we want to know the direct matrix for the entire space H, i.e.
\begin{equation}
  \bra{\alpha}\hat{A}_{\rm M}\ket{\beta} = \sum_{m,n\in{\rm M}} \bracket{\alpha}{m} (\widetilde{\Mat{A}}_{\rm M})_{mn} \bracket{n}{\beta}
\end{equation}
Hence the direct matrix of the operator $\hat{A}_{\rm M}$ is given by 
\begin{equation}
  \label{eq:MatSurProjection}
  \Mat{A}_{\rm M} =  \Mat{S}_{\rm HM} \tilde{\Mat{A}}_{\rm M} \Mat{S}_{\rm MH}
\end{equation}

\subsection{Projected Green's functions}

The central quantities both in DFT based transport calculations of molecular electronics 
devices and in quantum many-body theory are Green's functions (GF). The one-body GF is defined 
as the resolvent of the one-body Schr\"odinger equation~\cite{Economou::1970}:
\begin{equation}
  \hat{G}(z) (z+\mu-\hat{H}) = \hat{I}
\end{equation}
where z is complex, $\mu$ is the chemical potential, and $\hat{H}$ is the
Hamiltonian of the system. $\hat{G}(z)$ has poles at the eigen values $\epsilon_k$ 
of $\hat{H}$ for a finite system or a branch cut on the real axis at the 
energy bands for an infinite system. Its spectral representation in terms
of the eigen states $\ket{k}$ of $\hat{H}$ is given by:
\begin{equation}
  \hat{G}(z) = (z+\mu-\hat{H})^{-1} = \sum_k \frac{\ket{k}\bra{k}}{z+\mu-\epsilon_k}
\end{equation}
The GF operator projected onto subspace M is given by:
\begin{equation}
   \hat{G}_{\rm M}(z) = \hat{P}_{\rm M}\hat{G}(z)\hat{P}_{\rm M} 
   = \sum_{\alpha,\beta\in{M}} \ket{\alpha} (\widetilde{\bf G}_{\rm M}(z))_{\alpha\beta} \bra{\beta}
\end{equation}
Defining the GF of the isolated subspace M as
\begin{equation}
   \hat{g}_{\rm M}(z) = ((z+\mu)\hat{P}_{\rm M}-\hat{H}_{\rm M})^{-1}
\end{equation}
and the self-energy operator $\hat\Sigma_{\rm M}$ associated with 
the coupling of the subspace M to the rest of the world as
\begin{equation}
  \hat\Sigma_{\rm M}(z) = [\hat{g}_{\rm M}(z)]^{-1} - [\hat{G}_{\rm M}(z)]^{-1}
\end{equation}
it is possible to rewrite the projected GF as
\begin{equation}
  \label{eq:GMOp}
  \hat{G}_{\rm M}(z) = \left((z+\mu)\hat{P}_{\rm M}-\hat{H}_{\rm M}-\hat\Sigma_{\rm M}(z) \right)^{-1}
\end{equation}
The self-energy $\hat\Sigma_{\rm M}(z)$ is not to be confused with the one describing
electron-electron interactions in the many-body GF formalism. Note that in 
many-body physics in the context of the Anderson impurity model~\cite{Anderson:PR:1961}
$\hat\Sigma_{\rm M}(z)$ is often called \emph{hybridization function} 
and is denoted by $\hat\Delta_{\rm M}(z)$. 

One can easily write $\hat{\Sigma}_{\rm M}(z)$ in terms of the GF for 
the isolated (i.e. not coupled to M) complementary space $\bar{\rm M}$ defined by $\hat{P}_{\bar{\rm M}}$, 
$\hat{g}_{\bar{\rm M}}(z)=((z+\mu)\hat{P}_{\bar{\rm M}}-\hat{H}_{\bar{\rm M}})^{-1}$ as 
$\hat{\Sigma}_{\rm M}(z) = \hat{H}_{\rm M,\bar{M}} \, \hat{g}_{\bar{\rm M}}(z) \, \hat{H}_{\rm \bar{M},M}$
where $\hat{H}_{\rm M,\bar{M}}=\hat{P}_{\rm M}\hat{H}\hat{P}_{\bar{\rm M}}= (\hat{H}_{\rm \bar{M},M})^\dagger$.
In order to find the matrix representations $\Mat{G}_{\rm M}$ and $\widetilde{\Mat{G}}_{\rm M}$
of the projected GF $\hat{G}_{\rm M}(z)$, eq. (\ref{eq:GMOp}), is multiplied with the denominator 
of the r.h.s., the matrix elements are taken and the subspace identity $\hat{P}_{\rm M}$ is inserted
between the two factors of the l.h.s.:
\begin{eqnarray}
  \bra{\alpha}\hat{G}_{\rm M}(z) \sum_{\alpha^\prime,\beta^\prime\in{M}} \ket{\alpha^\prime} (\Mat{S}^{-1}_{\rm M})_{\alpha\beta^\prime} \bra{\beta^\prime}
  \left((z+\mu)\hat{P}_{\rm M}-\hat\Sigma_{\rm M}(z) \right)\ket{\beta} &=& \bra{\alpha}\hat{P}_{\rm M}\ket{\beta}
  \nonumber\\
\end{eqnarray}
Hence we find for the direct GF matrix
\begin{eqnarray}
  \Mat{G}_{\rm M}(z) &=& \Mat{S}_{\rm M} \left( (z+\mu)\Mat{S}_{\rm M} - \Mat{H}_{\rm M} - {\bf\Sigma}_{\rm M}(z) \right)^{-1} \Mat{S}_{\rm M}
  \nonumber\\
  &=& \left( (z+\mu)\Mat{S}_{\rm M}^{-1} - \widetilde{\bf H}_{\rm M} - \widetilde{\bf \Sigma}_{\rm M}(z) \right)^{-1}
\end{eqnarray}
and for the corresponding nuclear matrix
\begin{equation}
  \label{eq:GMTilde}
  \widetilde{\Mat{G}}_{\rm M}(z) = \left( (z+\mu)\Mat{S}_{\rm M} - \Mat{H}_{\rm M} - {\bf\Sigma}_{\rm M}(z) \right)^{-1}
\end{equation}

\subsection{Many-body Green's functions and Feynman diagrams in an atomic basis set}

The generalization of the one-body Green's function for an (effectively) non-interacting
system to the case of interacting electrons are the single-particle Green's function or 
single-particle propagators. 
The single-particle Matsubara GF~\cite{Mahan::2000} for atomic states $\alpha$ and $\alpha^\prime$ 
is defined as
\begin{equation}
  G_{\alpha\alpha^\prime}(\tau,\tau^\prime) = -\langle T_\tau[c_\alpha(\tau),c_{\alpha^\prime}^\dagger(\tau^\prime)] \rangle
\end{equation}
where $\tau$ is imaginary time and the creation and annihilation operators obey the generalized anti-commutation 
rules for non-orthogonal basis sets~\cite{Thygesen:PRB:2006}:
\begin{equation}
  \{ c_\alpha, c_\beta^\dagger \} = S_{\alpha\beta}
\end{equation}
The Fourier transform with respect to imaginary time $\tau$ yields the Matsubara GF
for imaginary frequencies (called Matsubara frequencies):
\begin{equation}
  G_{\alpha\alpha^\prime}(i\omega) = \int_0^\beta d\tau \, e^{i\omega\tau} G_{\alpha,\alpha^\prime}(\tau,0) 
\end{equation}
By analytic continuation to the real frequency axis one obtains the retarded single-particle 
GF $G_{\alpha\alpha^\prime}^{(+)}(\omega)\equiv{}G_{\alpha\alpha^\prime}(i\omega\rightarrow\omega+i\eta)$.

$G_{\alpha\alpha^\prime}$ defines the direct single-particle GF matrix ${\bf G}$. 
In the absence of interactions the single-particle GF matrix 
${\bf G}$ turns out to be equal to the one-body GF matrix defined as the resolvent of 
the one-body Schr\"odinger equation. 
Analogous to the one-body GF we can also define the nuclear matrix for the 
interacting single-particle GF as $\widetilde\Mat{G}=\Mat{S}^{-1}\Mat{G}\Mat{S}^{-1}$.

For the development of a diagrammatic expansion for the interacting GF in a non-orthogonal basis
in terms of the Coulomb interaction and the non-interacting GF, one has to either use the nuclear 
GF matrix in combination with the direct Coulomb interaction matrix, or the direct GF matrix in combination 
with the nuclear matrix of the Coulomb interaction~\cite{Thygesen:PRB:2006}. 
Here we will work with the nuclear matrix for the Green's functions and the direct matrix for the interactions.

The bare Coulomb interaction in an atomic basis set is given by:
\begin{equation}
  \mathcal{V}^{e-e} = \frac{1}{2}\sum_{\alpha,\alpha^\prime,\beta,\beta^\prime,\sigma,\sigma^\prime} 
  \widetilde{V}_{\alpha\beta;\alpha^\prime\beta^\prime} \, c^\dagger_{\alpha\sigma} c^\dagger_{\alpha^\prime\sigma^\prime} c_{\beta^\prime\sigma^\prime} c_{\beta\sigma} 
\end{equation}
where $\widetilde{V}_{\alpha\beta;\alpha^\prime\beta^\prime}$ is the nuclear matrix of the Coulomb interaction~\cite{Thygesen:PRB:2006},
i.e. $\widetilde{\Mat{V}}(1,2) = \Mat{S}(1)^{-1} \Mat{S}(2)^{-1} \Mat{V}(1,2) \Mat{S}(2)^{-1} \Mat{S}(1)^{-1}$
and the direct matrix elements are given by
\begin{eqnarray}
  V_{\alpha\beta;\alpha^\prime\beta^\prime} 
  = e^2 \int\int dr_1 dr_2 \frac{\phi_\alpha^\ast(r_1)\phi_{\beta}(r_1) \phi_{\alpha^\prime}^\ast(r_2)\phi_{\beta^\prime}(r_2)}{\|r_1-r_2\|}
\end{eqnarray}

\begin{figure}
  \includegraphics[width=\linewidth]{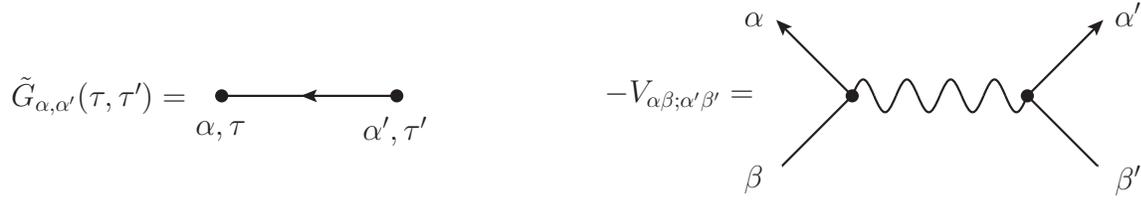}
  \caption{
    Feynman diagrams for the single-particle Green's function $\tilde{G}$ 
    and the bare Coulomb interaction $V_c$ in an atomic basis set.}
  \label{fig:feynman}
\end{figure}

The Feynman diagrams for the GF and the Coulomb interaction in an atomic basis set 
are shown in Fig.~\ref{fig:feynman}.

\subsection{DFT based transport calculations}
\label{sub:dft}

We consider the situation schematically depicted in Fig.~\ref{fig:scheme}(c). 
The central device region D containing a molecule is coupled to two electrodes 
L and R. This situation can be realized in a number of ways as shown in 
Figs.~\ref{fig:scheme}(a,b): (a) A molecule bridging the tips of a nanocontact
or (b) a molecule deposited on a metal substrate and coupled to an STM tip.
In addition to the molecule the device region D contains those parts of the two 
electrodes which are in close proximity to the molecule and whose electronic
structure is modified by the presence of the molecule and vice versa. 
In the case of the molecular bridge (a) the tips of the nanocontact are included 
in the device region while in the case of the molecule on the substrate (b), part 
of the surface and of the STM tip are included in the device region.

The electronic structure of the central device region is calculated ab initio
on the level of DFT in the Kohn-Sham (KS) framework, taking into account the 
coupling to the electrodes L and R. The Kohn-Sham Green's function of the device 
region D is given by:
\begin{equation}
  \widetilde{\Mat{G}}^0_{\rm D}(z) = ( (z+\mu)\Mat{S}_{\rm D}-\Mat{H}_{\rm D}^0 -\Mat{\Sigma}_{\rm L}(z) -\Mat{\Sigma}_{\rm R}(z) )^{-1}
\end{equation}
where $\Mat{H}_{\rm D}^0$ is the KS Hamiltonian of the device region which yields an
effective mean-field description of the electronic structure of the device region. 
$\Mat\Sigma_{\rm L}(z)$ and $\Mat\Sigma_{\rm R}(z)$ are the lead self-energies associated
with the coupling of the device region to the bulk electrodes. 

From the device GF the electronic density can easily be calculated  by
integration up to $\omega=0$ (corresponding to the chemical potential $\mu$)
\begin{equation}
  \label{eq:KSDensM}
  \widetilde{\Mat{D}}_{\rm D}^0 = -\Im \frac{1}{\pi} \int_{-\infty}^0 d\omega \, \widetilde{\Mat{G}}_{\rm D}^0(\omega+i\eta) 
\end{equation}
the density matrix yields a new KS Hamiltonian for the device region thus
closing the self-consistency cycle of the KS calculation. Hence we can self-consistently
calculate the electronic structure of the device region taking into account the coupling to the
electrodes (open system). 

In contrast to D, the electronic structure (Hamiltonian) of the electrodes
L and R, and hence the self-energies are kept fixed during the self-consistent
calculation of the electronic structure of D. Depending on the situation, 
different models for the bulk electrodes can be employed. One can 
for example choose nanowires~\cite{Taylor:PRB:2001}, 
embed the cluster into a perfect crystalline surface calculated ab initio~\cite{Brandbyge:PRB:2002}, 
or use so-called absorbing boundary conditions (ABC)~\cite{Baer:JCP:2004}.
Here we choose a tight-binding Bethe lattice model~\cite{Palacios::2005} 
with realistic tight-binding parameters obtained from DFT calculations~\cite{Papaconstantopoulos::1986}. 
The actual choice of the electrode model is not crucial for calculations as 
long as the bulk electrodes are far enough away from the central scattering 
region, i.e. the device region is chosen big enough and contains a sufficiently
big part of the electrodes~\cite{Jacob:JCP:2011}.

Once the KS calculation is converged the transport properties can be calculated
within the Landauer approach from the transmission function which is given by:
\begin{equation}
  \label{eq:transm}
  T^0(\omega) = \Tr[ \Mat\Gamma_{\rm L}(\omega) \widetilde{\Mat{G}}_{\rm D}^{0\dagger}(\omega) 
    \Mat\Gamma_{\rm R}(\omega) \widetilde{\Mat{G}}_{\rm D}^0(\omega) ]
\end{equation}
where $\Mat\Gamma_{\rm L}\equiv{}i(\Mat\Sigma_{\rm L}-\Mat\Sigma_{\rm L}^\dagger)$ and 
$\Mat\Gamma_{\rm R}\equiv{}i(\Mat\Sigma_{\rm R}-\Mat\Sigma_{\rm R}^\dagger)$ are the 
so-called coupling matrices which yield the broadening of the device 
region due to the coupling to the leads.
From the transmission function the current and conductance can be calculated
using the Landauer formula
\begin{eqnarray}
  \label{eq:current}
  I(V) = \frac{2e}{h} \int d\omega \, T^0(\omega)\, (f(\omega-\mu_L)-f(\omega-\mu_R)) 
\end{eqnarray}
where $\mu_{\rm L}$ and $\mu_{\rm R}$ are the electrochemical potentials of the left and right lead, 
respectively, defined by the applied bias voltage $eV=\mu_{\rm L}-\mu_{\rm R}$.
Note that in general the transmission function $T^0(\omega)$ also depends on the applied 
voltage $V$, i.e. $T^0=T^0(\omega,V)$, and actually has to be calculated out of equilibrium
by combining the KSDFT with the NEGF~\cite{Taylor:PRB:2001a,Palacios:PRB:2002,Brandbyge:PRB:2002}.
However, within the mean-field like KSDFT based NEGF approach the transmission is often
not so strongly voltage dependent, and hence current and conductance can be approximated 
well by the equilibrium transmission $T^0(\omega,0)$ at least for sufficiently small bias 
voltages. 

In the typical situation of an STM setup (Fig.~\ref{fig:scheme}(b)), most of the applied
bias voltage $V$ will drop near the sharp STM tip, i.e. the electrochemical potential
of the substrate remains fixed to the equilibrium one $\mu_{\rm sub}=\mu$ while that
of the STM tip changes with the bias $\mu_{\rm tip}=\mu+eV$. 
The differential conductance for low bias at zero temperature is then directly given by
the transmission function:
\begin{equation}
  \label{eq:conductance_asymm}
  G(V) = \frac{\partial{I}}{\partial{V}} 
  =  \frac{2e}{h} \times \frac{\partial}{\partial{V}} \int_{0}^{eV} d\omega \, T^0(\omega) 
  = \frac{2e^2}{h} \times T^0(eV)
\end{equation}
In contrast, for the situation of a molecule coupled symmetrically to two leads
(Fig.~\ref{fig:scheme}(a)), the voltage will drop more or less symmetrically
across the junction, i.e. $\mu_L=\mu-eV/2$ and $\mu_R=\mu+eV/2$. Hence for the
conductance we obtain now
\begin{equation}
  \label{eq:conductance_symm}
  G(V) =  \frac{2e}{h} \times \frac{\partial}{\partial{V}} \int_{-eV/2}^{+eV/2} d\omega \, T^0(\omega) 
  = \frac{e^2}{h} \left[ T^0\left(\frac{eV}{2}\right) + T^0\left(-\frac{eV}{2}\right) \right]
\end{equation}
In a more general situation where the coupling is neither completely symmetric nor completely
asymmetric, more sophisticated modelling of the electrostatics or even a KS-NEGF calculation
would be necessary in order to find the actual voltage drop.

\subsection{Projection onto the correlated subspace: Anderson impurity model}
\label{sub:aim}

Next we have to identify the strongly correlated subspace C. Usually C will be 
formed by the open $d$- or $f$-shells of transition metal atoms. However, also
molecular orbitals of purely organic molecules such as C$_{60}$ or carbon nanotubes 
weakly coupled to electrodes can show strong correlations if the effective
interaction in these levels is big in comparison with the broadening due to the
coupling to the leads. Our approach is completely general in this respect.

From now on we assume that the orbitals $\phi$ forming the subspace C are mutually
orthogonal (but not necessarily orthogonal to the other orbitals in the device
region). This can always be achieved by simple L\"owdin 
orthogonalization of subspace C. However, note that often the orbitals spanning
C are already mutually orthogonal. For example in the case of the atomic orbitals
forming the open $d$- or $f$-shell of a transition metal atom, or in the case
of molecular orbitals which are the eigenstates of the KS Hamiltonian of the 
molecule and hence by construction are orthogonal.
In order to account for the strong correlations in subspace C an effective
Coulomb interaction term 
\begin{equation}
  \hat{\mathcal{V}}^{e-e}_{\rm C} = \frac{1}{2} \sum_{{ijkl}\atop{\sigma\sigma^\prime}} U_{ik;jl} \, c_{i\sigma}^\dagger c_{j\sigma^\prime}^\dagger c_{l\sigma^\prime} c_{k\sigma}
\end{equation}
is added acting on the orbitals in C. Note that $U_{ik;jl}$ is not the bare Coulomb interaction
but an effective interaction which is usually much lower than the bare one due to screening
processes by the conduction electrons. In the next section it is shown how to calculate 
$U_{ik;jl}$ ab initio from the DFT electronic structure.
The full many-body Hamiltonian of the strongly interacting subspace C now reads:
\begin{equation}
  \hat{\mathcal{H}}_{\rm C} = \hat{\mathcal{H}}_{\rm C}^0 + \hat{\mathcal{V}}^{e-e}_{\rm C}
\end{equation}
where the one-body part $\hat{\mathcal{H}}_{\rm C}^0 = \sum_{i,j,\sigma} \bra{\phi_i} \hat{H}_{\rm C}^0 \ket{\phi_j} c_{i\sigma}^\dagger c_{j\sigma}$ 
is given by projection of the KS Hamiltonian $\hat{H}_{\rm D}^0$ onto C.
However, since the Coulomb interaction has been taken into account already
on a mean-field level in the Kohn-Sham Hamiltonian, we also need to subtract a double-counting correction (DCC) 
term:
\begin{equation}
  \hat{H}_{\rm C}^0 = \hat{P}_{\rm C} \hat{H}_{\rm D}^0 \hat{P}_{\rm C} - \hat{V}_{\rm C}^{dc} 
\end{equation}
Unfortunately, the DCC term $\hat{V}_{\rm C}^{dc}$ is not exactly known for DFT, and several approximation 
schemes are used in practice~\cite{Karolak:JESRP:2010}. 
Here the so-called atomic limit or fully localized limit (FLL) is employed~\cite{Czyzyk:PRB:1994}, 
but generalized to the case of an anisotropic Coulomb repulsion $U_{ii;jj}$~\cite{Jacob:PRB:2013}:
\begin{equation}
  \label{eq:FLL}
  (V^{dc}_{\rm C})_{ii} = \sum_{j} U_{ii;jj} \, \left(n_{j}-\frac{1}{2M_{\rm C}}\right) - J_{\rm H} \frac{ N_{\rm C} - 1}{2}
\end{equation}
where $n_j=\langle c_j^\dagger c_j\rangle$ is the electronic occupation of orbital $\phi_j$,
$M_{\rm C}$ is the dimension of subspace C,
$J_{\rm H}$ is the Hund's rule coupling given by the orbital-averaged exchange matrix elements $U_{ij;ji}$,
and $N_{\rm C}=\sum_{j\in{\rm C}} n_j$ is the total electronic occupation of subspace C.

According to (\ref{eq:GMTilde}) the self-energy (a.k.a. the hybridization function)
associated with the coupling of C to the rest of the system
is given by 
\begin{equation}
  \label{eq:}
  \Delta_{\rm C}(\omega) = (\omega+\mu)\Mat{1}_{\rm C} -\Mat{H}_{\rm C}^0 - [\widetilde{\Mat{G}}_{\rm C}^0(\omega)]^{-1}
\end{equation}
where the projected GF of the correlated subspace $\rm{}C\subset{}D$ can be calculated from the 
device GF according to (\ref{eq:MatSubProjection}) as 
\begin{equation}
  \widetilde{\Mat{G}}_{\rm C}(\omega) = \Mat{S}_{\rm CD} \widetilde{\Mat{G}}_{\rm D}(\omega) \Mat{S}_{\rm DC}
\end{equation}
As customary in many-body physics we will call $\Delta_{\rm C}(\omega)$ the hybridization
function from now on.
The many-body Hamiltonian $\hat{\mathcal{H}}_{\rm C}$ of subspace C together with the hybridization function 
$\Delta_{\rm C}(\omega)$ define a multi-orbital Anderson impurity model (AIM). 
Solution of the AIM yields the self-energy $\Sigma_{\rm C}(\omega)$ describing the 
strong electronic correlations within the C subspace which is fed-back to the DFT
calculation in order to obtain electronic spectra and transport properties of the
molecular device (see Sec.~\ref{sub:feedback}).

\subsection{Computation of the effective interaction in the correlated subspace}
\label{sub:crpa}

The effective interaction $\hat{\mathcal{V}}_{\rm C}^{e-e}$ between the electrons in the correlated subspace C
is not the bare Coulomb interaction because of screening processes by formation of electron-hole (e-h) 
pairs in the rest of the system. Therefore the screened Coulomb matrix elements $U_{ik;jl}$ are
considerably lower than the bare Coulomb interaction $V_{ik;jl}$. The screening of the bare interaction by 
formation of e-h pairs can be calculated within the so-called Random Phase Approximation (RPA) 
(see e.g. the book by Mahan~\cite{Mahan::2000} or any other textbook on quantum many-body theory).
However, screening of the electrons within the C subspace will already be taken into account by 
the impurity solver. Hence the contribution of the impurity subspace C to the screening needs
to be subtracted out. By doing so one arrives at the so-called constrained Random Phase Approximation 
(cRPA)~\cite{Aryasetiawan:PRB:2006}.

In order to calculate the effective screened interaction $U_{ik;jl}$ of subspace C within cRPA we first
define the so-called polarizability region P in which screening processes due to formation of e-h pairs 
are taken into account for calculating the screened interaction.
P comprises the strongly correlated subspace C and a sufficient portion of the surrounding atoms
of subspace C  as is schematically indicated in Fig.~\ref{fig:scheme}. In principle, the whole D region
could be chosen as P. However, in practice this is often not feasible because of computational limitations
if the device region is reasonably big. Also as it turns out the screening of the correlated subspace C by 
the surrounding conduction electrons is relatively localized due to the usually localized nature of the 
strongly correlated orbitals making up C. 

\begin{figure}
  \includegraphics[width=\linewidth]{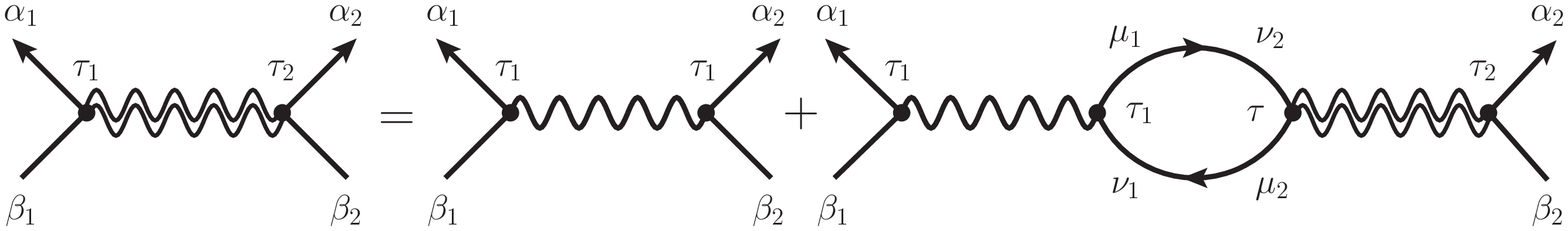}
  \caption{
    Dyson equation for RPA screened interaction for atomic basis set.
    Wiggly lines correspond to the bare Coulomb interaction $V$, 
    double wiggly lines to the RPA screened interaction $W$.
  }
  \label{fig:rpa}
\end{figure}

Within RPA the screened interaction $W$ is given by the Dyson equation shown diagrammatically in Fig. \ref{fig:rpa}
which in an atomic orbital basis set can be written algebraically as
\begin{eqnarray}
  \label{eq:dyson-rpa}
  \lefteqn{-W_{\alpha_1\beta_1;\alpha_2\beta_2}(\tau_1,\tau_2) = -V_{\alpha_1\beta_1;\alpha_2\beta_2} \times \delta(\tau_1-\tau_2) }
  \nonumber\\
  && -\sum_{\mu_1\nu_1\mu_2\nu_2} V_{\alpha_1\beta_1;\mu_1\nu_1} \int_0^\beta d\tau\, (\widetilde{\Mat\Pi}_{\rm P})_{\mu_1\nu_1;\mu_2\nu_2}(\tau_1,\tau) 
  \, W_{\mu_2\nu_2;\alpha_2\beta_2}(\tau,\tau_2)
\end{eqnarray}
For the screening of the bare Coulomb interaction $V$ only screening processes within region P are taken 
into account. Hence we have to calculate the polarizability (i.e. the bubble diagram in Fig.~\ref{fig:rpa}) 
projected onto the P region:
\begin{equation}
  (\widetilde{\Mat\Pi}_{\rm P})_{\alpha\beta;\alpha^\prime\beta^\prime}(\tau,\tau^\prime) 
  = \sum_\sigma (\widetilde{\Mat{G}}^0_{\rm P})^\sigma_{\beta^\prime\alpha}(\tau^\prime,\tau) \, 
  (\widetilde{\Mat{G}}^0_{\rm P})^\sigma_{\beta\alpha^\prime}(\tau,\tau^\prime)
\end{equation}
where the projected GF for the P region can be obtained from the device GF according to (\ref{eq:MatSubProjection}) as 
\begin{equation}
  \widetilde{\Mat{G}}_{\rm P} = \Mat{S}_{\rm P}^{-1} \, \Mat{S}_{\rm PD} \, \widetilde{\Mat{G}}_{\rm D} \, \Mat{S}_{\rm DP} \, \Mat{S}_{\rm P}^{-1}
\end{equation}

For a stationary Hamiltonian we can replace the two times in the screened interaction 
and polarizability by time differences: $\Pi(\tau_1,\tau_2)\rightarrow\Pi(\tau_1-\tau_2)$ and
$W(\tau_1,\tau_2)\rightarrow W(\tau_1-\tau_2)$. 
Hence (by setting $\tau_2=0$ and after some renaming), we can write the Dyson equation 
for the RPA screened interaction as:
\begin{eqnarray}
  \label{eq:dyson-rpa2}
  \lefteqn{W_{\alpha_1\beta_1;\alpha_2\beta_2}(\tau) = V_{\alpha_1\beta_1;\alpha_2\beta_2} \times \delta(\tau)  }
  \nonumber\\
  && +\sum_{\mu_1\nu_1\mu_2\nu_2} V_{\alpha_1\beta_1;\mu_1\nu_1} \int_0^\beta d\tau^\prime \, (\widetilde{\Mat\Pi}_{\rm P})_{\mu_1\nu_1;\mu_2\nu_2}(\tau-\tau^\prime) 
  W_{\mu_2\nu_2;\alpha_2\beta_2}(\tau^\prime)
\end{eqnarray}

Here we will only consider the static limit of the screened interaction, i.e.
$W^0\equiv W(\omega=0) = \int d\tau\,W(\tau)$. 
Because of the $\beta$-periodicity of $\Pi(\tau)$ we also have
$\int_0^\beta d\tau\,\Pi(\tau-\tau^\prime)=\int_0^\beta d\tau\,\Pi(\tau)\equiv\Pi^0$.
Hence we obtain the following Dyson equation for the {\it static} screened interaction 
$W^0$:
\begin{eqnarray}
  \label{eq:dyson-rpa3}
  W^0_{\alpha_1\beta_1;\alpha_2\beta_2} = V_{\alpha_1\beta_1;\alpha_2\beta_2} + \sum_{\mu_1\nu_1\mu_2\nu_2} V_{\alpha_1\beta_1;\mu_1\nu_1} (\widetilde{\Pi}_{\rm P}^0)_{\mu_1\nu_1;\mu_2\nu_2} W^0_{\mu_2\nu_2;\alpha_2\beta_2}
\end{eqnarray}

The {\it static} Polarizability $\Pi^0$ is now found easily by integrating a Green's function product over the frequency domain:
\begin{eqnarray}
  (\widetilde{\Mat\Pi}^0_{\rm P})_{\mu_1\nu_1;\mu_2\nu_2} 
  &&\equiv \int_0^\beta d\tau\, (\widetilde{\Mat\Pi}_{\rm P})_{\mu_1\nu_1;\mu_2\nu_2}(\tau)
  = \int_0^\beta d\tau\, \sum_\sigma (\widetilde{\Mat{G}}_{\rm P}^0)_{\nu_2\mu_1}^\sigma(-\tau)\,(\widetilde{\Mat{G}}_{\rm P}^0)_{\nu_1\mu_2}^\sigma(\tau) 
  \nonumber\\
  &&= \frac{1}{\beta} \sum_{i\omega_n} \sum_\sigma (\widetilde{\Mat{G}}_{\rm P}^0)_{\nu_2\mu_1}^\sigma(i\omega_n)\,(\widetilde{\Mat{G}}_{\rm P}^0)_{\nu_1\mu_2}^\sigma(i\omega_n) 
  \nonumber\\
  &&\stackrel{\displaystyle\longrightarrow}{\scriptscriptstyle \beta\rightarrow\infty}
  \frac{1}{2\pi} \int_{-\infty}^\infty d\omega\, \sum_\sigma (\widetilde{\Mat{G}}_{\rm P}^0)_{\nu_2\mu_1}^\sigma(i\omega)\,(\widetilde{\Mat{G}}_{\rm P}^0)_{\nu_1\mu_2}^\sigma(i\omega) 
\end{eqnarray}
where in the last step we have taken the zero temperature limit ($\beta\rightarrow\infty$)
rendering the discrete Matsubara frequencies continuous.

We now define superindices $\underline{I}:=(i_1,i_2)$ in order to rewrite
the Dyson equation in form of a matrix equation. 
Hence we have $\mathbf{W}^0=(W^0_{\underline{I},\underline{J}})$ etc.,
and the Dyson equation can be written in matrix form as:
\begin{eqnarray}
  \mathbf{W}^0 &=& \mathbf{V} + \mathbf{V}\, \widetilde{\Mat{\Pi}}_{\rm P}^0 \, \mathbf{W}^0
\end{eqnarray}
Solving for the static screened interaction $\mathbf{W}^0$ we find:
\begin{equation}
  \label{eq:rpa}
  \mathbf{W}^0 = \left( \mathbf{1} - \mathbf{V}\, \widetilde{\Mat{\Pi}}_{\rm P}^0\right)^{-1} \mathbf{V}
\end{equation}

\begin{figure}
  \includegraphics[width=\linewidth]{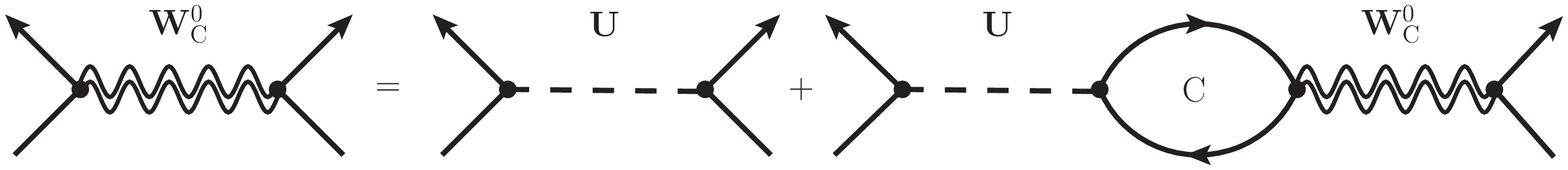}
  \caption{\label{fig:crpa}
    Dyson equation for fully screened RPA interaction $\mathbf{W}_{\rm C}^0$ of subspace C 
    in terms of effective interaction $\mathbf{U}$. Orbital indexes have been suppressed here.
  }
\end{figure}

Projection to the correlated subspace C then yields the RPA screened interaction for the correlated
electrons $\Mat{W}^0_{\rm C}$.
However, since the screening within the correlated subspace will already be taken into
account by the impurity solver in a more or less exact way, the screening of the correlated electrons
by themselves has to be subtracted out in order to obtain the effective interaction $\mathbf{U}$.
Hence the effective interaction $\mathbf{U}$ is the partially screened interaction that results in the 
fully RPA screened interaction $\mathbf{W}_{\rm C}^0$ when taking into account only the polarizability $\widetilde{\mathbf{\Pi}}_{\rm C}$ 
within the C subspace. The corresponding Dyson equation is shown diagrammatically in Fig.~\ref{fig:crpa}.
Solving for the effective interaction $\mathbf{U}$ we arrive at the following 
``unscreening'' equation~\cite{Aryasetiawan:PRB:2004,Haule:PRB:2010,Vaugier:PRB:2012}
computing $\mathbf{U}$ from $\mathbf{W}_{\rm C}^0$:
\begin{equation}
  \label{eq:U}
  \Mat{U} = \Mat{W}_{\rm C}^0 \, (\Mat{1}_{\rm C} + \widetilde{\Mat\Pi}^0_{\rm C} \Mat{W}_{\rm C}^0 )^{-1}
\end{equation}
In order to determine the screening within subspace C, we have to calculate the polarizability
corresponding to subspace C. 
\begin{equation}
  \widetilde{\Mat\Pi}_{\rm C}^0 = \frac{1}{2\pi} \int_{-\infty}^\infty d\omega\, \sum_\sigma 
  (\widetilde{\Mat{G}}_{\rm C})_{\nu_2\mu_1}^\sigma(i\omega)\, (\widetilde{\Mat{G}}_{\rm C})_{\nu_1\mu_2}^\sigma(i\omega)
\end{equation}
where $ \widetilde{\Mat{G}}_{\rm C}(i\omega) = \Mat{S}_{\rm CD} \widetilde{\Mat{G}}_{\rm D}(i\omega) \Mat{S}_{\rm DC}$. 

It is important to realize that $\widetilde{\Mat\Pi}_{\rm C}^0$ and $\widetilde{\Mat{G}}_{\rm C}$ are not just
submatrices of the corresponding bigger matrices $\widetilde{\Mat\Pi}_{\rm P}^0$ and $\widetilde{\Mat{G}}_{\rm P}$
in the P region of the device due to the overlap between the subspaces. Neglecting this detail can result
in serious errors in the computation of the effective Coulomb interaction $\Mat{U}$: Due to the numerical
instability of eq.~(\ref{eq:U}) small inaccuracies in computing $\widetilde{\Mat\Pi}_{\rm C}^0$ can result
in large errors and even in completely unphysical effective interactions. The numerical instability of 
eq.~(\ref{eq:U}) can be seen by rewriting it as
\begin{equation}
  \Mat{U} = \left(\left[\Mat{W}_{\rm C}^0\right]^{-1} + \widetilde{\Mat\Pi}^0_{\rm C} \right)^{-1}
\end{equation}
As the fully screened interaction $\Mat{W}_{\rm C}^0$ is usually quite small (compared to the bare Coulomb interaction)
and positive, $[\Mat{W}_{\rm C}^0]^{-1}$ is big and positive. On the other hand the screening of the correlated electrons
by themselves is usually strong, and therefore $\widetilde{\Mat\Pi}^0_{\rm C}$ is big and negative. Hence in order
to obtain $\Mat{U}$ we are basically subtracting two relatively big numbers and inverting the resulting small number 
so that relatively small errors in calculating $\Mat{W}_{\rm C}^0$ or $\widetilde{\Mat\Pi}^0_{\rm C}$ can result in quite 
large errors for $\Mat{U}$. 
It should be noted here that in the case of a semiconducting or insulating substrate or host material, 
as well as in the case of insulating compounds this issue is less problematic since then at low energies
around the Fermi level, the two subspaces are completely decoupled, leading to weaker ``self-screening'' 
of the correlated electrons, and hence smaller numbers for $[\Mat{W}_{\rm C}^0]^{-1}$ and $\widetilde{\Mat\Pi}^0_{\rm C}$.

However, in the case of a metallic host or substrate considered here, it is thus crucial to correctly perform the projections 
of the different quantities involved in the calculation of the effective interaction (Green's functions, polarizability) 
to the P and C subspaces in order to reliably calculate $\Mat{U}$. 
Also the usual neglect of certain product basis states in the computation of the screened 
interaction~\cite{Aryasetiawan:RPP:1998} might be problematic in this context.
One way to stabilize the numerical evaluation of (\ref{eq:U}) is to decouple the correlated subspace from
the rest of the system both in the calculation of $\Mat{W}$ and of $\Mat{U}$ as proposed by Miyake {\it et al}.~\cite{Miyake:PRB:2009}. 
This way the self-screening of the correlated electrons is reduced considerably, leading to smaller values of 
$[\Mat{W}_{\rm C}^0]^{-1}$ and $\widetilde{\Mat\Pi}^0_{\rm C}$, and thus enhancing the numerical stability.
However, this can lead to far too high matrix elements for the direct Coulomb interaction as will be shown in Sec.~\ref{sec:results}. 
Apparently, ``mixed propagators'' between the correlated subspace and the rest of the system (which vanish when the subspaces are 
decoupled) can be quite important for the screening of the effective interaction.

\subsection{Solution of the Anderson impurity model: One-Crossing Approximation}
\label{sub:oca}

Since the interaction $U_{ijkl}$ is strong in comparison with the single-particle broadening 
(given by the imaginary part of $\hat\Delta_{\rm C}(\omega)$), the AIM problem cannot
be solved by standard perturbation theory in the Coulomb interaction. Instead more advanced
many-body methods usually starting from an exact diagonalization of the full impurity Hamiltonian 
$\hat{H}_{\rm C}$ have to be employed in order to properly take into account the strong correlations
within subspace C. Here I use the One-Crossing Approximation (OCA)~\cite{Haule:PRB:2001}
which is an improvement over the Non-Crossing Approximation (NCA)~\cite{Grewe:PRB:1981,Kuramoto:ZPB:1983,Coleman:PRB:1984}.
However, it should be emphasized that the methodology presented so far can in principle be combined 
with any other method for solving the AIM, as e.g. continuous time Quantum Monte-Carlo (CTQMC)~\cite{Gull:RMP:2011}, 
or numerical renormalization group (NRG)~\cite{Bulla:RMP:2008}, or the Lanczos diagonalization scheme~\cite{Ishida:PRB:2012}. 

One advantage of OCA over other schemes is that spectral data can be calculated directly on the
real frequency axis. Hence in contrast to the numerically exact CTQMC, for example, it does not 
suffer from artifacts introduced by numerical analytic continuation of the spectra from the Matsubara axis 
to the real axis. Also spurious features in the spectra coming from the approximation of the infinite and continuous
conduction electron bath in the Anderson model by a finite and discrete one as in direct diagonalization 
schemes such as Lanczos, are not a problem for OCA since the bath is not truncated or discretized.
On the other hand, in contrast to the basically exact but computationally very demanding NRG, OCA can 
actually be applied to realistic Anderson models of $3d$- and $4f$-impurities with 5 and 7 impurity-levels,
respectively.

However, being an approximate method, OCA also suffers from some deficiencies that one should 
be aware of. First, as in the case of the simpler NCA, spurious non-Fermi liquid behaviour is obtained in 
the zero-temperature limit, resulting in artifacts in the spectral density for low temperatures.
While in NCA these artifacts already appear below $T_K$, in OCA the critical temperature below
which the artifacts appear is significantly lower (1-2 orders below $T_K$). 
Another problem of NCA and OCA is the violation of certain sum rules especially in the case of
multi-orbital Anderson models that lead to errors in the high frequency expansion of the electronic 
self-energy~\cite{Ruegg:PRB:2013}. Again, these errors are much less pronounced in OCA than in NCA. 

The basic idea of both NCA and OCA methods is to treat the coupling of the correlated subspace
C to the rest of the system given by the hybridization function $\Delta_{\rm C}(\omega)$
as a perturbation to the dynamics within the subspace induced by the strong electron-electron
interactions which is treated exactly.
Hence the starting point is an exact diagonalization of the many-body Hamiltonian of 
the correlated subspace:
\begin{equation}
  \hat{\mathcal{H}}_{\rm C} = \sum_m E_m \ket{m}\bra{m}
\end{equation}
where $\ket{m}$ are the many-body eigenstates of $\mathcal{H}_{\rm C}$ and $E_m$ the corresponding eigen-energies.

It is now convenient to represent the  many-body  eigenstates of $\ket{m}$, 
in terms of auxiliary fields or {\it pseudo-particles} (PPs)
$\hat{a}_m,\hat{a}_m^\dagger$ which obey (anti-)commutation rules depending on the 
number of electrons represented by the corresponding many-body state $\ket{m}$.
The physical electron operators $c_{i\sigma}$, $c_{i\sigma}^\dagger$ 
are related to the PP operators by:
\begin{equation}
  \label{eq:PPtoReal}
  c_{i\sigma} = \sum_{m,n} F_{i\sigma}^{mn} a_m^\dagger a_n
\end{equation}
where $F_{i\sigma}^{mn}$ are the matrix elements of the electron annihilation operator
with the many-body eigenstates of C: $F_{i\sigma}^{mn}=\bra{m}c_{i\sigma}\ket{n}$.
Since the PPs obey (anti-)commutation rules a diagrammatic expansion of PP propagators
in terms of the coupling to the rest of the system is possible. 
The full PP propagator corresponding to a many-body state $\ket{m}$ is then given by
\begin{equation}
  G_m(\omega) = \frac{1}{\omega-\lambda-E_m-\Sigma_m(\omega)}
  \label{eq:ppgf}
\end{equation}
where $\Sigma_m(\omega)$ is the PP self-energy describing the dynamic interaction with the other PPs
induced by the hybridization with the rest of the system (bath). 

\begin{figure}
  \includegraphics[width=\linewidth]{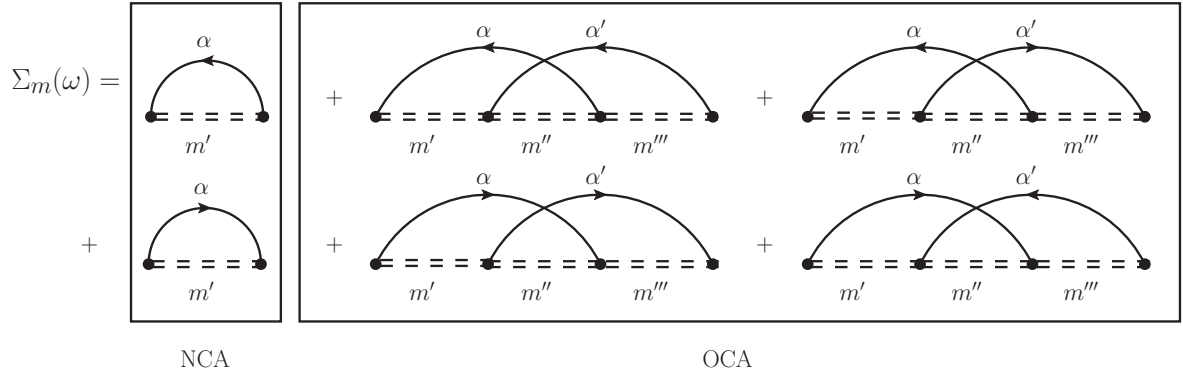}
  \caption{Diagrams for pseudo-particle self-energies in NCA and OCA approximations.
    full lines correspond to conduction electron propagators coupled to impurity levels
    $\alpha$, double dashed lines to full pseudo-particle propagators.}
  \label{fig:nca+oca}
\end{figure}

NCA consists in an infinite resummation of self-energy diagrams where conduction electron lines 
do not cross (hence the name). These are the diagrams shown in the left box in Fig. \ref{fig:nca+oca} 
for a certain PP $m$ representing a many-body state of $N$ electrons. The NCA diagrams describe 
processes where a single electron (hole) jumps from the bath to subspace C and back thereby 
temporarily creating a PP with N+1 (N-1) electrons. 
Hence the NCA self-energy is given by a convolution of the hybridization function 
$\Delta_{\rm C}(\omega)$ with the PP propagators $G_{m^\p}$ of the PPs $m^\p$ coupled to $m$.
OCA additionally takes into account diagrams where two bath electron lines cross 
as shown in the right box of Fig.~\ref{fig:nca+oca}. The algebraic expressions 
for the OCA self-energies involve double convolutions of two hybridization functions 
with three PP propagators. The exact algebraic expressions can be found in the 
literature~\cite{Haule:PRB:2001,Haule:PRB:2010}.
Since the self-energy of a PP $m$ depends on the dressed propagators
of the other PPs $m^\p$ that interact via $\mathcal{V}_{\rm hyb}$ with $m$,
the NCA/OCA equations have to be solved self-consistently. 

Once the NCA/OCA equations have been solved, the real electronic
quantities can be calculated from the PP propagators by expanding 
the real electron operators in terms of PP operators by (\ref{eq:PPtoReal}).
Within NCA, the real electron spectral
function is obtained from the PP spectral functions as
\begin{eqnarray}
  \label{eq:rhod}
  \rho_{i\sigma}(\omega) &=& \frac{1}{\langle{Q}\rangle} \sum_{mm^\prime} \int d\varepsilon\,
  e^{-\beta\varepsilon}[1+e^{-\beta\omega}]
  |F_{i\sigma}^{mm^\prime}|^2 \, A_m(\varepsilon) \, A_{m^\prime}(\omega+\varepsilon)
\end{eqnarray}
where $A_m(\omega)=-\Im\,G_m(\omega)/\pi$ is the PP spectral function for
PP $m$ and $Q$ is the PP charge which is obtained by integration of the
PP spectral functions. Again. in OCA the expression for calculating 
the electronic density $\rho_{i\sigma}(\omega)$ is more complicated, involving
double convolutions of PP spectral functions.
From the electron spectral density $\rho_{i\sigma}(\omega)$ being the imaginary part (modulo $\pi$) 
of the electron Green's function $G_{i\sigma}(\omega)$ we can calculate the real part 
of $G_{i\sigma}(\omega)$ by Kramers-Kronig.
Finally, from the GF $\hat{G}_{\rm C}(\omega)$ the electronic self-energy
describing the dynamic correlations within C is obtained by
$\hat{\Sigma}_{\rm C}(\omega)=[\hat{G}_{\rm C}^0(\omega)]^{-1}-[\hat{G}_{\rm C}(\omega)]^{-1}$
where $\hat{G}_{\rm C}^0(\omega)=((\omega+\mu)\hat{P}_{\rm C}-\hat{H}_{\rm C}^0-\hat\Delta_{\rm C}(\omega))^{-1}$
is the bare propagator of subspace C.
For a more detailed account of the NCA, OCA and other methods based on
a hybridization expansion of atomic states see e.g. Refs.~\cite{Kotliar:RMP:2006,Haule:PRB:2010}.

\subsection{Feedback of the self-energy: correlated electronic structure and transport properties}
\label{sub:feedback}

Once we have solved the Anderson impurity model for the strongly interacting subspace C coupled
to the rest of the system, we obtain the electronic self-energy describing the strong dynamic 
correlations within subspace C:
\begin{equation}
  \label{eq:SigmaC}
  \hat\Sigma_{\rm C}(\omega) = \sum_{i,j\in{\rm C}} \ket{i} \left[\Mat\Sigma_{\rm C}(\omega)\right]_{ij} \bra{j}
\end{equation}
Note that $\widetilde{\Mat\Sigma}_{\rm C}=\Mat\Sigma_{\rm C}$ since we have assumed the basis to be
orthonormal within subspace C.
This self-energy is now fed back to the DFT part in order to obtain the correlated electronic structure and transport 
properties of the system. More specifically, we obtain the correlated device GF
\begin{equation}
  \hat{G}_{\rm D}(\omega) = \left( [\hat{G}_{\rm D}^0(\omega)]^{-1} - [\hat\Sigma_{\rm C}(\omega) - \hat{V}_{\rm C}^{dc}] \right)^{-1}
\end{equation}
where $\hat{V}_{\rm C}^{dc}$ is the DCC operator which like $\hat\Sigma_{\rm C}(\omega)$ only acts on C.
According to (\ref{eq:MatSurProjection}) the corresponding nuclear matrix of the device GF is 
\begin{equation}
  \widetilde{\Mat{G}}_{\rm D}(\omega) = \left( [\widetilde{\Mat{G}}_{\rm D}^0(\omega)]^{-1} 
  - \Mat{S}_{\rm DC} [\Mat\Sigma_{\rm C}(\omega) - \Mat{V}_{\rm C}^{dc}] \Mat{S}_{\rm CD} \right)^{-1}
\end{equation}
where the overlap matrices $\Mat{S}_{\rm DC}$ and $\Mat{S}_{\rm CD}$ sandwiching 
$\Mat\Sigma_{\rm C}(\omega) - \Mat{V}_{\rm C}^{dc}$ account for the overlap between 
the correlated subspace C and the rest of the system (see eq.~\ref{eq:MatSurProjection}).

From the correlated device GF $\widetilde\Mat{G}_{\rm D}(\omega)$ we can calculate the correlated electronic density
analogously to (\ref{eq:KSDensM}) by integration of $\widetilde\Mat{G}_{\rm D}(\omega)$ up to 0 energy:
\begin{equation}
  \widetilde{\Mat{D}}_{\rm D} = -\Im \frac{1}{\pi} \int_{-\infty}^0 d\omega \, \widetilde{\Mat{G}}_{\rm D}(\omega+i\eta) 
\end{equation}
From the correlated density in turn a new KS Hamiltonian for the device region can be calculated, from which a new 
correlated density is obtained and so forth until self-consistency is reached. Hence we can calculate the
effect of the correlation within the C subspace onto the charge distribution of the device region. 
This part corresponds to the so-called ``charge self-consistency'' loop within the DFT+DMFT scheme~\cite{Pourovskii:PRB:2007}.

Following Meir-Wingreen~\cite{Meir:PRL:1992}, the low-bias transport properties can be 
obtained in complete analogy to the case of KS-DFT transport eqs.~(\ref{eq:transm}-\ref{eq:conductance_symm}) 
even in the presence of strong correlations from the \emph{correlated} transmission function
\begin{equation}
  \label{eq:corrtransm}
  T(\omega) = \Tr[ \Mat\Gamma_{\rm L}(\omega) \widetilde{\Mat{G}}_{\rm D}^\dagger(\omega) \Mat\Gamma_{\rm R}(\omega) \widetilde{\Mat{G}}_{\rm D}(\omega) ]
\end{equation}
Note that the strong correlations giving rise e.g. to the Kondo effect are actually contained
in $T(\omega)$ via the correlated GF $\widetilde{\Mat{G}}_{\rm D}(\omega)$.

In the next section we will see that the Fano-Kondo lineshapes measured by STM spectroscopy 
of magnetic atoms and molecules on metal substrates can indeed be reproduced by calculating
the conductance from the (zero-bias) transmission function. This is due to the fact that
the Kondo effect is a low-energy phenomenon, i.e. the Kondo peak is observed for very small 
bias voltages so that finite-bias effects only play a minor role. 
For the description of actual non-equilibrium phenomena the formalism has to be generalized
to include the effect of finite bias voltages. As shown by Meir and Wingreen in their landmark
papers~\cite{Meir:PRL:1992} this can be achieved by generalization of the formalism to the Keldysh 
contour. However, in this case the Anderson impurity problem has to be solved out of equilibrium 
which is computationally extremely demanding. So far it has only been achieved in the context of
the single-level AIM~\cite{Meir:PRL:1993,Wingreen:PRB:1994,Cohen:PRL:2014}, but not for realistic cases.

\section{Results: Co adatom at the Cu(001) surface}
\label{sec:results}

\begin{figure}
  \includegraphics[width=\linewidth]{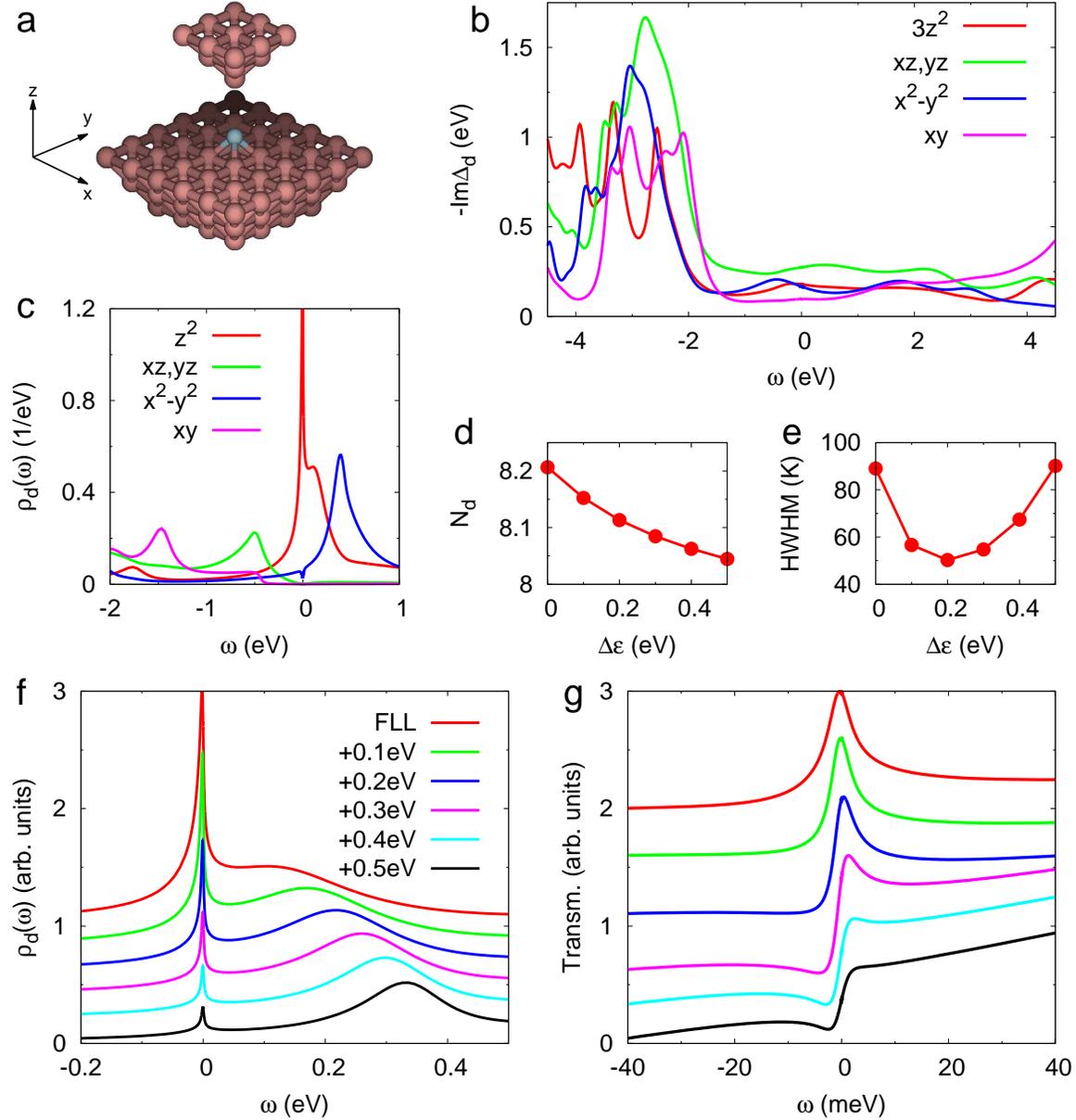}
  \caption{
    \label{fig:results}
    Results for Co adatom at Cu(001) surface. 
    (a) Atomic structure of device part. The Co adatom is shown in grey.
    (b) Orbitally resolved imaginary part of hybridization functions for Co $3d$-shell.
    (c) Orbitally resolved OCA spectral functions for Co $3d$-shell at $T\sim10$K. 
    (d) Total occupation of Co $3d$-shell as a function of energy shift $\Delta\varepsilon$.
    (e) Half-width of Kondo feature in $z^2$ spectral function as a function
    of the total shift $\Delta\varepsilon$ of Co $3d$-levels with respect to
    energy levels given by FLL DCC.
    (f) Spectral functions of Co $z^2$-orbital for different energy shifts $\Delta\varepsilon$
    at $T\sim10$K.
    (g) DFT+OCA transmission functions for different energy shifts $\Delta\varepsilon$ 
    (line colours as in (f)).
  }
\end{figure}

Now the developed methodology is applied to the case of a Co adatom 
deposited on the Cu(001) surface. This system is an ideal testbed for
the theory as it has been measured extensively in the recent 
past~\cite{Knorr:PRL:2002,Wahl:PRL:2004,Neel:PRL:2007,Neel:PRL:2008,Uchihashi:PRB:2008,Vitali:PRL:2008}.
Fig.~\ref{fig:results}(a) shows the atomic structure of the device 
region. The device contains the Co atom on three layers of the 
Cu(001) surface and an STM tip consisting of a Cu pyramid 
grown in the (001) direction. The Co atom and its four nearest 
neighbour Cu atoms have been relaxed with Gaussian09~\cite{G09}
using the local spin density approximation (LSDA) and the LANL2DZ 
double-zeta valence plus outer core electron basis set
with core pseudo potentials~\cite{Hay:JCP:1985} 
while the rest of the device atoms have been kept fixed. 
The interlayer and intralayer distances for the fixed Cu atoms are those of a 
perfect Cu surface taken from Ref.~\cite{DaSilva:PRB:2004}.
In good agreement with Ref.~\cite{Pick:PRB:2003,Vitali:PRL:2008}, I find that 
the Co atom relaxes at a height of about 1.5\r{A} above the four nearest neighbour 
Cu atoms while these in turn are pushed by 0.1\r{A} into the substrate.

\begin{table}
  \begin{flushright}
  \begin{tabular}{l|ccccccc}
    {}             & $N_{d}$ & $z^2$ & $xz$ & $yz$ & $x^2-y^2$ & $xy$ & $S_d$ \\
    \hline
    LSDA           &   8.13 & 1.59  & 1.66 & 1.66 & 1.33      & 1.89 & 0.82  \\  
    LDA            &   8.25 & 1.66  & 1.63 & 1.63 & 1.52      & 1.81 &  -    \\
    OCA            &   8.26 & 1.34  & 1.94 & 1.94 & 1.08      & 1.97 & 0.86  \\
    OCA ($+0.4$eV) &   8.15 & 1.11  & 1.96 & 1.96 & 1.06      & 1.97 & 0.94
  \end{tabular}
  \end{flushright}
  \caption{\label{tab:occupations}
    Total and orbital resolved occupations and spin of Co $3d$-shell within DFT 
    on the level of LSDA and LDA and DFT+OCA calculations. In the last line we
    show the DFT+OCA results for the Co $3d$-levels $\epsilon_d$ shifted by 
    0.4eV upwards in energy with respect to the FLL.
  }
\end{table}

Using ANT.G and the LANL2MB minimal basis set including valence and outer core electrons 
with pseudo potentials~\cite{Hay:JCP:1985} the electronic and magnetic structure structure 
of the device coupled to the tip and substrate electrodes is calculated within DFT on the
level of LSDA. The Co atom is essentially in a $4s^13d^8$ configuration with the two holes 
in the $3d$-shell giving rise to an a approximate spin-1 state of the Co atom 
(see Tab.~\ref{tab:occupations}) again in good agreement with~\cite{Pick:PRB:2003}.
LSDA basically predicts a mixed valence situation for all the orbitals with the 
individual occupations around 1.6 with the exception of the $xy$-orbital which
is nearly full. 

From the LSDA electronic structure the effective Coulomb interaction $U_{ij;kl}$
for the Co $3d$-shell is calculated as described in Sec.~\ref{sub:crpa}. For the
P region we take into account substrate atoms up to the 3rd nearest neighbour,
i.e. the 9 Cu atoms closest to the Co adatom. The change in $U$ from taking
into account 2nd nearest neighbours to 3rd nearest neighbours is about 2\%.
For 4th nearest neighbours (18 atoms in total) the super matrices in the RPA 
equation (\ref{eq:rpa}) become too big (linear matrix dimension $234^2=54756$) to be 
handled. 

Tab.~\ref{tab:coulomb} shows the matrix elements of the effective Coulomb interaction, 
namely the direct Coulomb repulsion matrix elements (density-density interaction) $U_{ii;kk}$ 
and the exchange interaction matrix elements (Hund's rule coupling) $U_{ik;ki}$.
The average density-density interaction is $\bar{U}=4.14$~eV. It is
strongly screened by the conduction electrons, resulting in a reduction of over 80\% compared 
to the bare value of 22.9~eV for the Co $3d$-shell. On the other hand, the Hund's 
rule coupling is much less affected by the screening: it is only reduced by about 10\% 
from its bare value of 0.85~eV, resulting in an average Hund's coupling of $J_H=0.77$~eV.   
Note that the inter-orbital Coulomb repulsion ($U_{ii;kk}$ for $i\ne{k}$) is related 
to the average intra-orbital Coulomb repulsion for both orbitals and the Hund's rule 
coupling via $U_{ii;kk}=(U_{ii;ii}+U_{kk;kk})/2-2U_{ik;ki}$. 

Both density-density interactions and Hund's rule coupling are somewhat anisotropic 
(i.e. orbital-dependent). The intra-orbital Coulomb repulsion $U_{ii;ii}$ deviates only 
by up to 0.17eV or by to 3\% from its mean value of $U=5.4$eV. 
The variation is stronger for the inter-orbital Coulomb repulsion $U_{ii;kk}$, deviating 
by up to 0.43eV or by up to 11\% from its mean value of $U^\prime=3.85$eV. 
The Hund's rule coupling  $U_{ik;ki}$ deviates even stronger by up to 0.29eV 
or by up to 38\% from its mean value of $J_H=0.77eV$.
It is worth noting at this point that the complete decoupling of the correlated subspace 
from the rest of the system as proposed in Ref.~\cite{Miyake:PRB:2009} in order to achieve 
a stable computation of the effective interaction in the case of ``entangled bands'' 
produces a much higher density-density interaction of about 12eV. Apparently the screening
effects of ``mixed propagators'' between the correlated subspace and the rest of the system
are actually quite important and cannot be neglected.

\begin{table}
  \begin{flushright}
    \begin{tabular}{r|ccccc|ccccc}
      {}        & \multicolumn{5}{c}{density-density interaction (eV)} & \multicolumn{4}{|c}{Hund's coupling (eV)} \\
      {}        & $z^2$ & $xz$ & $yz$ & $x^2-y^2$ & $xy$ & $z^2$ & $xz$ & $yz$ & $x^2-y^2$ \\
      \hline
      $z^2$     & 5.38  & 4.27 & 4.27 & 3.45 & 3.46   \\
      $xz$      & 4.27  & 5.56 & 3.86 & 3.73 & 3.74   & 0.60 \\ 
      $yz$      & 4.27  & 3.86 & 5.56 & 3.73 & 3.74   & 0.60 & 0.83 \\
      $x^2-y^2$ & 3.45  & 3.73 & 3.73 & 5.23 & 4.28   & 0.94 & 0.82 & 0.82 \\
      $xy$      & 3.46  & 3.74 & 3.74 & 4.28 & 5.26   & 0.92 & 0.83 & 0.83 & 0.48
    \end{tabular}
    \caption{\label{tab:coulomb}
      Direct Coulomb repulsion matrix elements $U_{ii;kk}$ (density-density interaction)
      and exchange matrix elements $U_{ik;ki}$ (Hund's rule coupling) of effective Coulomb 
      interaction for Co $3d$-shell.
    }
  \end{flushright}
\end{table}

Next, the electronic structure of the system is calculated for the paramagnetic
case on the level of the local density-approximation (LDA) in order to obtain the 
KS energy levels of the Co $3d$-shell $\epsilon_d^0$ and hybridization functions 
$\Delta_d(\omega)$ in the absence of spin-polarization.
Fig.~\ref{fig:results}(b) shows the imaginary parts of the hybridization functions
$\Delta_d(\omega)$ for each of the Co $3d$-orbitals which yields the (dynamic) broadening
of the orbitals due to the coupling to the substrate. We see that the broadening near the 
Fermi level is basically featureless indicating coupling to the delocalized Cu $4s$-states
of the substrate. As can be seen the degenerate $xz$- and $yz$-levels couple most strongly 
to the these states. Because of their shape these two orbitals couple very well to the 
$4s$-states of the four Cu atoms directly underneath the Co adatom. 
On the other hand, the coupling of the $xy$-orbital to the substrate is the
weakest for all five orbitals since the direct coupling to the underneath 
Cu atoms is strongly suppressed due to symmetry reasons.
The coupling of the $z^2$- and the $x^2-y^2$-orbitals to the substrate is intermediate 
between these two cases. 
At negative energies, the coupling to the localized Cu $3d$-states of the substrate leads to 
strong peaks in the hybridization functions at energies between -5 and -2~eV. 
Less pronounced peaks at positive energies above 4eV indicate coupling to the Cu $4p$-orbitals
of the substrate.

The bare energies $\epsilon_d$ of the Co $3d$-levels constituting the impurity shell 
in the Anderson impurity model are obtained from their KS energies $\epsilon_d^0=\hat{P}_d\hat{H}^0\hat{P}_d$ 
corrected by a DCC term, as explained earlier in Sec.~\ref{sub:aim}. The so-called
FLL generalized to an anisotropic (i.e. orbital-dependent) density-density 
interaction $U_{ii;kk}$ is employed (\ref{eq:FLL}).
The values for the direct Coulomb repulsion are the ones shown in Tab.~\ref{tab:coulomb}. For 
the Hund's rule coupling the orbital averaged exchange interaction is taken, i.e. $J_H=0.77$eV.

The Anderson impurity problem presented by the interacting Co $3d$-shell coupled to the 
substrate is now solved within OCA as described in Sec.~\ref{sub:oca}. For the effective
Coulomb interaction of the Co $3d$-shell we take into account the density-density interactions 
$U_{ii;kk}$ as well as the exchange interactions $U_{ik;ki}$ as given in Tab.~\ref{tab:coulomb}.
At the energy levels for the Co $3d$-orbitals given by the FLL-DCC (\ref{eq:FLL}), the
total occupancy for the Co $3d$-shell is about 8.2 electrons similar to the ones
of the LDA and LSDA calculations (see Tab.~\ref{tab:occupations}). 
However, the individual occupancies of the $3d$-orbitals are now quite different 
from the DFT ones, namely they are now closer to integer occupancies, as opposed to
the mixed-valence situations obtained in the DFT calculations. In particular, the 
$x^2-y^2$-orbital is now basically half-filled, and the $xz$-, $yz$- and $xy$-orbitals
are nearly full now. The $z^2$-orbital is now also closer to half-filling than before 
but still has strong charge fluctuations (occupancy$\sim1.3$). Similar to LSDA, the 
spin of the Co $3d$-shell is found to be $S_d\sim$0.87, close to a spin-1 configuration. 

In Fig.~\ref{fig:results}(c) the calculated spectral functions of the Co $3d$-orbitals
$\rho_d(\omega)$ (at $T\sim10K$) are shown. We can see a very strong Kondo peak at the Fermi level in the 
$z^2$-orbital. The upper Hubbard peak is here quite close to the Kondo peak at the Fermi level
due to the strong charge fluctuations. This orbital is still quite close to a mixed-valence situation.
The $x^2-y^2$-orbital despite being half-filled and thus bearing a spin-1/2 does not yield 
a Kondo peak. We are dealing here essentially with a so-called underscreened 
Kondo effect~\cite{Nozieres:JPhys:1980,Coleman:PRB:2003,Posazhennikova:PRB:2007}: 
Despite the relatively similar hybridization of the $z^2$-channel and the $x^2-y^2$-channel, 
the Kondo temperature $T_{K,z^2}$ of the $z^2$-channel is much higher than that of the 
$x^2-y^2$-channel, $T_{K,x^2-y^2}$, due to its stronger charge fluctuations. Hence at finite 
temperature $T$ with $T_{K,x^2-y^2}<T<T_{K,z^2}$, only the spin-1/2 in the $z^2$-channel is 
Kondo-screened, while the spin-1/2 in the $x^2-y^2$ channel remains unscreened. 

The half-width of the Kondo peak is about 90~K, in very good agreement with the experimentally 
observed values~\cite{Knorr:PRL:2002,Neel:PRL:2007,Uchihashi:PRB:2008}.
However, the lineshape of the calculated transmission function 
(red curve in Fig.~\ref{fig:results}(g)) is rather peak-like, different from the experimentally 
observed asymmetric Fano-lineshaphes. As the DCC for DFT is not exactly known and eq.~(\ref{eq:FLL}) 
is only an approximation, we now shift the Co $3d$-levels upwards in energy by an amount 
$\Delta\varepsilon$ thus emptying the Co $3d$-shell as can be seen in Fig.~\ref{fig:results}(d).
Emptying the Co $3d$-shell mainly lowers the occupancy of the $z^2$-orbital reducing the charge 
fluctuations for that orbital, while the occupancies of the other orbitals are quite stable.
Fig.~\ref{fig:results}(f) shows the effect of shifting the Co $3d$-levels and the concomitant
reduction of charge fluctuations on the the spectral function of the $z^2$-orbital (at $T\sim10$K): 
As the $z^2$-orbital is emptied, its occupation approaches 1, the Kondo peak becomes smaller, and 
the upper Hubbard peak moves away from the Fermi level. The width of the Kondo peak decreases
at first and then starts to grow again for shifts $\ge0.2$eV, as can be seen in Fig.~\ref{fig:results}(e). 
Note that the non-monotonic behaviour of the width of the Kondo peak is actually a finite temperature 
effect: As the actual Kondo temperature decreases with decreasing charge fluctuations, the Kondo 
peak in the finite temperature spectra (here $T\sim10$K) does not attain its full (zero-temperature) 
height anymore. As the height of the (finite-$T$) Kondo peak decreases, its width starts to grow again 
at some point. Hence the half-width of the Kondo peak measured at some finite temperature really only 
yields an \emph{apparent} Kondo temperature.

Fig.~\ref{fig:results}(g) shows the effect of shifting the Co $3d$-levels on the low-energy transmission 
spectra. As said above, for the $3d$-levels at the values given by the FLL-DCC the transmission function 
near zero energy is rather peak-like, unlike the ones observed experimentally. But when shifting the 
$3d$-levels upwards in energy the lineshapes become more asymmetric Fano-like. Good agreement between theoretical
and experimental Fano-lineshapes is achieved for shifts between 0.4eV to 0.5eV. In this regime the half-width 
of the Kondo peak is between 67K and 86K, in good agreement with the experimentally observed ones between 
70K and 100K for the Co on Cu(001) system~\cite{Knorr:PRL:2002,Neel:PRL:2007,Uchihashi:PRB:2008}.  
These results are quite different from those obtained recently for the case of Co on Cu(111) with a similar
approach~\cite{Surer:PRB:2012} where all the Co $3d$-orbitals give rise to Kondo-like resonances at the
Fermi level. The reason could be the altogether quite different geometric situation at the (111) surface
leading to a decidedly different symmetry and occupancy for the Co $3d$-shell.

\section{Conclusions}
\label{sec:conclusions}

In conclusion, an ab initio methodology has been developed for describing the impact of 
strong electronic correlations on the electronic structure and transport properties 
of nanoscale devices. Starting from the DFT electronic structure of an embedded nanoscale 
device, an Anderson impurity model is constructed by projection of the Kohn-Sham Hamiltonian 
onto the correlated subspace. The effective Coulomb interaction $U$ for the correlated 
subspace (impurity) is calculated ab initio from the DFT electronic structure by making 
use of the constrained RPA approach. The solution of the Anderson impurity model yields 
the dynamic correlations originating from strong interactions within the correlated subspace 
in form of a self-energy which is fed back to the DFT calculation in order to obtain 
the correlated electronic structure and transport properties.

The methodology has been tested for the case of a single Co adatom on Cu(001) substrate. 
On a qualitative level the results are in good agreement with experiments: A Fano-Kondo
feature with the width in good agreement with experiments is obtained in the calculated 
low-energy tunnelling spectra. However, the lineshape of the Fano-Kondo feature is not 
correctly reproduced at the energies for the Co $3d$-levels given by the double-counting 
correction. Only when shifting the Co $3d$-levels slightly upwards in energy good agreement 
with the experimentally observed lineshapes is achieved. 
It is a well known problem of DFT+U and DFT+DMFT approaches that the double-counting correction
is not exactly known and in general does not yield the correct position (and thus charge) of the 
correlated levels. Nevertheless, the so-called fully-localized limit employed here, is actually
not too far off as only moderate shifts are necessary to achieve good quantitative agreement 
with experiments. Importantly, the physics is actually not affected by the shifting of the Co 
$3d$-levels: Independent of the shift (in that energy range) the Co $3d$-shell constitutes
essentially a spin-1 system that experiences an underscreened Kondo effect. The shifting only 
affects the weight of the Kondo peak by lowering the charge fluctuations in the Kondo-screened 
orbital.

Hence the developed methodology is capable of qualitative predictions of  
strong correlation phenomena. But accurate quantitative predictions for example 
of Kondo temperatures and the exact shapes of Fano-Kondo features are 
difficult as these are dependent on the exact occupancy of the correlated 
subspace which cannot be calculated accurately because of the approximate 
nature of the double-counting correction in our approach. 
One possibility to overcome these difficulties is to make use of the GW approach
instead of DFT for the description of the weakly interacting part of the system,
similar to the GW+DMFT approach for strongly correlated 
materials~\cite{Biermann:PRL:2003,Taranto:PRB:2013,Hansmann:PRB:2013}
since for GW the double-counting correction term is exactly known.

\ack
I would like to thank Juanjo Palacios and Maria Soriano for fruitful discussions 
about projections in non-orthogonal basis sets, and Silke Biermann and Kristjan 
Haule for stimulating discussions on the calculation of the effective interaction U. 
I am also grateful to Hardy Gross for his support and encouragement.

\section*{References}

\bibliographystyle{unsrt}
\bibliography{myrefs}

\begin{thebibliography}{100}

\bibitem{Heath:PT:2003}
J.~R. Heath and A.~R. Ratner.
\newblock Molecular electronics.
\newblock {\em Physics Today}, 56:43, 2003.

\bibitem{Joachim:PNAS:2005}
Christian Joachim and Mark~A. Ratner.
\newblock Molecular electronics: Some views on transport junctions and beyond.
\newblock {\em Proceedings of the National Academy of Sciences of the United
  States of America}, 102:8801, 2005.

\bibitem{Cuniberti::2005}
G.~Cunibert, G.~Fags, and K.~Richter.
\newblock {\em Introducing Molecular Electronics}.
\newblock Springer, Berlin, 2005.

\bibitem{Cuevas::2010}
J.~C. Cuevas and E.~Scheer.
\newblock {\em Molecular Electronics}.
\newblock World Scientific, Singapore, 2010.

\bibitem{Aviram:CPL:1974}
A.~Aviram and M.~A. Ratner.
\newblock Molecular rectifiers.
\newblock {\em Chem. Phys. Lett.}, 29:277, 1974.

\bibitem{Elbing:PNAS:2005}
M.~Elbing, R.~Ochs, M.~Koentopp, M.~Fischer, C.~von Hänisch, F.~Weigend,
  F.~Evers, H.~B. Weber, and M.~Mayor.
\newblock A single-molecule diode.
\newblock {\em Proc. Natl. Acad. Sci.}, 102:8815, 2005.

\bibitem{Tans:N:1998}
S.~J. Tans, A.~R.~M. Verscheren, and C.~Dekker.
\newblock Room-temperature transistor based on single carbon nanotube.
\newblock {\em Nature}, 393:49, 1998.

\bibitem{Martel:APL:1998}
R.~Martel, T.~Schmidt, H.~R. Shea, T.~Hertel, and Ph. Avouris.
\newblock Single- and multi-wall carbon nanotube field-effect transistors.
\newblock {\em Applied Physics Letters}, 73:2447, 1998.

\bibitem{Zutic:RMP:2004}
I.~\v{Z}uti\'{c}, J.~Fabian, and S.~Das~Sarma.
\newblock Spintronics: Fundamentals and applications.
\newblock {\em Rev. Mod. Phys.}, 76:323, 2004.

\bibitem{Bogani:NM:2008}
L.~Bogani and W.~Wernsdorfer.
\newblock Molecular spintronics using single-molecule magnets.
\newblock {\em Nature Materials}, 7:179, 2008.

\bibitem{Gatteschi::2006}
D.~Gatteschi, R.~Sessoli, and J.~Villain.
\newblock {\em {Molecular Nanomagnets}}.
\newblock Oxford University Press, 2006.

\bibitem{Agrait:PR:2003}
N.~Agra\"it, A.~L. Yegati, and J.~M. van Ruitenbeek.
\newblock Quantum properties of atomic-sized conductors.
\newblock {\em Physics Reports}, 377:81, 2003.

\bibitem{Liang:N:2001}
W.~Liang, M.~Bockrath, D.~Bozovic, J.~H. Hafner, M.~Tinkham, and H.~Park.
\newblock Fabry-Perot interference in a nanotube electron waveguide.
\newblock {\em Nature}, 411:665, 2001.

\bibitem{Kong:PRL:2001}
J.~Kong, E.~Yenilmez, T.~W. Tombler, W.~Kim, H.~Dai, R.~B. Laughlin, L.~Liu,
  C.~S. Jayanthi, and S.~Y. Wu.
\newblock Quantum interference and ballistic transmission in nanotube electron
  waveguides.
\newblock {\em Phys. Rev. Lett.}, 87:106801, 2001.

\bibitem{Roch:N:2008}
N.~Roch, S.~Florens, V.~Bouchiat, W.~Wernsdorfer, and F.~Balestro.
\newblock Quantum phase transitions in a single-molecule quantum dot.
\newblock {\em Nature}, 453:633, 2008.

\bibitem{Palacios:PRL:2003}
J.~J. Palacios, A.~J. P\'erez-Jim\'enez, E.~Louis, E.~SanFabi\'an, and J.~A.
  Verg\'es.
\newblock First-principles phase-coherent transport in metallic nanotubes with
  realistic contacts.
\newblock {\em Phys. Rev. Lett.}, 90:106801, 2003.

\bibitem{Palacios:PRB:2008}
J.~J. Palacios, P.~Tarakeshwar, and Dae~M. Kim.
\newblock Metal contacts in carbon nanotube field-effect transistors: Beyond
  the Schottky barrier paradigm.
\newblock {\em Phys. Rev. B}, 77:113403, 2008.

\bibitem{Schull:PRL:2009}
G.~Schull, T.~Frederiksen, M.~Brandbyge, and R.~Berndt.
\newblock Passing current through touching molecules.
\newblock {\em Phys. Rev. Lett.}, 103:206803, 2009.

\bibitem{Taylor:PRB:2001a}
J.~Taylor, H.~Guo, and J.~Wang.
\newblock Ab initio modeling of quantum transport properties of molecular
  electronic devices.
\newblock {\em Phys. Rev. B}, 63:245407, 2001.

\bibitem{Palacios:PRB:2002}
J.~J. Palacios, A.~J. P\'erez-Jim\'enez, E.~Louis, E.~SanFabi\'an, and J.~A.
  Verg\'es.
\newblock First-principles approach to electrical transport in atomic-scale
  nanostructures.
\newblock {\em Phys. Rev. B}, 66:035322, 2002.

\bibitem{Brandbyge:PRB:2002}
M.~Brandbyge, J.~L. Mozos, P.~Ordej\'on, J.~Taylor, and K.~Stokbro.
\newblock Density functional method for nonequilibrium electron transport.
\newblock {\em Phys. Rev. B}, 65:165401, 2002.

\bibitem{Rocha:PRB:2006}
A.~R. Rocha, V.~M. Garcia-Suarez, S.~Bailey, C.~Lambert, J.~Ferrer, and
  S.~Sanvito.
\newblock Spin and molecular electronics in atomically generated orbital
  landscapes.
\newblock {\em Phys. Rev. B}, 73:085414, 2006.

\bibitem{Mehrez:PRB:2002}
H.~Mehrez, A.~Wlasenko, B.~Larade, J.~Taylor, P.~Gr\"utter, and H.~Guo.
\newblock I-V characteristics and differential conductance fluctuations of Au
  nanowires.
\newblock {\em Phys. Rev. B}, 65:195419, 2002.

\bibitem{DiVentra:PRL:2000}
M.~Di~Ventra, S.~T. Pantelides, and N.~D. Lang.
\newblock First-principles calculation of transport properties of a molecular
  device.
\newblock {\em Phys. Rev. Lett.}, 84:979, 2000.

\bibitem{Varga:PRL:2007}
K.~Varga and S.~T. Pantelides.
\newblock Quantum transport in molecules and nanotube devices.
\newblock {\em Phys. Rev. Lett.}, 98:076804, 2007.

\bibitem{Lindsay:AM:2007}
S.~M. Lindsay and M.~A. Ratner.
\newblock Molecular transport junctions: Clearing mists.
\newblock {\em Advanced Materials}, 19:23, 2007.

\bibitem{Kondo:PTP:1964}
J.~Kondo.
\newblock Resistance minimum in dilute magnetic alloys.
\newblock {\em Prog. Theor. Phys.}, 32:37, 1964.

\bibitem{Hewson::1997}
A.~C. Hewson.
\newblock {\em The Kondo problem to heavy fermions}.
\newblock Cambridge University Press, Cambridge, 1997.

\bibitem{Madhavan:S:1998}
V.~Madhavan, W.~Chen, T.~Jamneala, M.~F. Crommie, and N.~S. Wingreen.
\newblock {Tunneling into a Single Magnetic Atom: Spectroscopic Evidence of the
  Kondo Resonance}.
\newblock {\em Science}, 280:567, 1998.

\bibitem{Li:PRL:1998}
J.~Li, W.-D. Schneider, R.~Berndt, and B.~Delley.
\newblock Kondo scattering observed at a single magnetic impurity.
\newblock {\em Phys. Rev. Lett.}, 80:2893, 1998.

\bibitem{Park:N:2002}
J.~Park, A.N. Pasupathy, J.~I. Goldsmith, C.~Chang, Y.~Yaish, J.~R. Petta,
  M.~Rinkoski, J.~P. Sethna, H.~Abru\~na, P.~L. McEuen, and D.~C. Ralph.
\newblock Coulomb blockade and Kondo effect in single-atom transistors.
\newblock {\em Nature}, 417:722, 2002.

\bibitem{Liang:N:2002}
W.~Liang, M.~P. Shores, M.~Bockrath, J.~R. Long, and H.~Park.
\newblock Kondo resonance in a single-molecule transistor.
\newblock {\em Nature}, 417:729, 2002.

\bibitem{Zhao:S:2005}
A.~Zhao, Q.~Li, L.~Chen, H.~Xiang, W.~Wang, S.~Pan, B.~Wang, X.~Xiao, J.~Yang,
  J.~G. Hou, and Q.~Zhu.
\newblock Controlling the Kondo effect of an adsorbed magnetic ion through its
  chemical bonding.
\newblock {\em Science}, 309:1542, 2005.

\bibitem{Yu:PRL:2005}
L.~H. Yu, Z.~K. Keane, J.~W. Ciszek, L.~Cheng, J.~M. Tour, T.~Baruah, M.~R.
  Pederson, and D.~Natelson.
\newblock Kondo resonances and anomalous gate dependence in the electrical
  conductivity of single-molecule transistors.
\newblock {\em Phys. Rev. Lett.}, 95:256803, 2005.

\bibitem{Iancu:NL:2006}
V.~Iancu, A.~Deshpande, and S.-W. Hla.
\newblock Manipulating Kondo temperature via single molecule switching.
\newblock {\em Nano Letters}, 6:820, 2006.

\bibitem{Fu:PRL:2007}
Y.-S. Fu, S.-H. Ji, X.~Chen, X.-C. Ma, R.~Wu, C.-C. Wang, W.-H. Duan, X.-H.
  Qiu, B.~Sun, P.~Zhang, J.-F. Jia, and Q.-K. Xue.
\newblock Manipulating the Kondo resonance through quantum size effects.
\newblock {\em Phys. Rev. Lett.}, 99:256601, 2007.

\bibitem{Calvo:N:2009}
M.~R. Calvo, J.~Fern\'andez-Rossier, J.~J. Palacios, D.~Jacob, D.~Natelson, and
  C.~Untiedt.
\newblock The Kondo effect in ferromagnetic atomic contacts.
\newblock {\em Nature}, 358:1150, 2009.

\bibitem{Franke:S:2011}
K.~J. Franke, G.~Schulze, and J.~I. Pacual.
\newblock Competition of superconducting phenomena and Kondo screening at the
  nanoscale.
\newblock {\em Science}, 332:940, 2011.

\bibitem{Minamitani:PRL:2012}
E.~Minamitani, N.~Tsukahara, D.~Matsunaka, Y.~Kim, N.~Takagi, and M.~Kawai.
\newblock Symmetry-driven novel Kondo effect in a molecule.
\newblock {\em Phys. Rev. Lett.}, 109:086602, 2012.

\bibitem{Kuegel:NL:2014}
J.~K\"ugel, M.~Karolak, J.~Senkpiel, P.-J. Hsu, G.~Sangiovanni, and M.~Bode.
\newblock Relevance of hybridization and filling of 3d orbitals for the Kondo
  effect in transition metal phthalocyanines.
\newblock {\em Nano Lett.}, 14:3895, 2014.

\bibitem{Nygard:N:2000}
J.~Nygard, D.~H. Cobden, and P.~E. Lindelof.
\newblock Kondo physics in carbon nanotubes.
\newblock {\em Nature}, 408:342, 2000.

\bibitem{Yu:NL:2004}
L.~H. Yu and D.~Natelson.
\newblock The Kondo effect in C$_60$ single-molecule transistors.
\newblock {\em Nano Letters}, 4:79, 2004.

\bibitem{Jarillo-Herrero:N:2005}
P.~Jarillo-Herrero, J.~Kong, H.~S.~J. van~der Zant, C.~Dekker, L.~P.
  Kouwenhoven, and S.~De~Franceschi.
\newblock Orbital Kondo effect in carbon nanotubes.
\newblock {\em Nature}, 434:484, 2005.

\bibitem{Parks:PRL:2007}
J.~J. Parks, A.~R. Champagne, G.~R. Hutchison, S.~Flores-Torres, H.~D.
  Abru\~na, and D.~C. Ralph.
\newblock Tuning the Kondo effect with a mechanically controllable break
  junction.
\newblock {\em Phys. Rev. Lett.}, 99:026601, 2007.

\bibitem{Roch:PRL:2009}
N.~Roch, S.~Florens, T.~A. Costi, W.~Wernsdorfer, and F.~Balestro.
\newblock Observation of the underscreened Kondo effect in a molecular
  transistor.
\newblock {\em Phys. Rev. Lett.}, 103:197202, 2009.

\bibitem{Bergfield:PRL:2012}
J.~P. Bergfield, Z.-F. Liu, K.~Burke, and C.~A. Stafford.
\newblock Bethe ansatz approach to the Kondo effect within density-functional
  theory.
\newblock {\em Phys. Rev. Lett.}, 108:066801, 2012.

\bibitem{Mera:PRL:2010}
H.~Mera and Y.~M. Niquet.
\newblock Are Kohn-Sham conductances accurate?
\newblock {\em Phys. Rev. Lett.}, 105:216408, 2010.

\bibitem{Stefanucci:PRB:2004}
G.~Stefanucci and C.-O. Almbladh.
\newblock Time-dependent partition-free approach in resonant tunneling systems.
\newblock {\em Phys. Rev. B}, 69:195318, 2004.

\bibitem{DiVentra:JoPCM:2004}
M.~Di~Ventra and T.~N. Todorov.
\newblock Transport in nanoscale systems: the microcanonical versus
  grand-canonical picture.
\newblock {\em Journal of Physics: Condensed Matter}, 16:8025, 2004.

\bibitem{Darancet:PRB:2007}
P.~Darancet, A.~Ferretti, D.~Mayou, and V.~Olevano.
\newblock Ab initio GW electron-electron interaction effects in quantum
  transport.
\newblock {\em Phys. Rev. B}, 75:075102, 2007.

\bibitem{Thygesen:JCP:2007}
K.~S. Thygesen and A.~Rubio.
\newblock Non-equilibrium GW approach to quantum transport in nanoscale
  contacts.
\newblock {\em J. Chem. Phys.}, 126:091101, 2007.

\bibitem{Darancet:NL:2012}
P.~Darancet, J.~R. Widawsky, H.~J. Choi, L.~Venkataraman, and J.~B. Neaton.
\newblock Quantitative current-voltage characteristics in molecular junctions
  from first principles.
\newblock {\em Nano Letters}, 12:6250, 2012.
\newblock PMID: 23167709.

\bibitem{Stefanucci:PRL:2011}
G.~Stefanucci and S.~Kurth.
\newblock Towards a description of the Kondo effect using time-dependent
  density-functional theory.
\newblock {\em Phys. Rev. Lett.}, 107:216401, 2011.

\bibitem{Turkowski:JoPCM:2014}
V.~Turkowski and T.~S. Rahman.
\newblock Nonadiabatic time-dependent spin-density functional theory for
  strongly correlated systems.
\newblock {\em Journal of Physics: Condensed Matter}, 26:022201, 2014.

\bibitem{Jacob:PRL:2009}
D.~Jacob, K.~Haule, and G.~Kotliar.
\newblock Kondo effect and conductance of nanocontacts with magnetic
  impurities.
\newblock {\em Phys. Rev. Lett.}, 103:016803, 2009.

\bibitem{Jacob:PRB:2010}
D.~Jacob and G.~Kotliar.
\newblock Orbital selective and tunable Kondo effect of magnetic adatoms on
  graphene: Correlated electronic structure calculations.
\newblock {\em Phys. Rev. B}, 82:085423, 2010.

\bibitem{Jacob:PRB:2010a}
D.~Jacob, K.~Haule, and G.~Kotliar.
\newblock Dynamical mean-field theory for molecular electronics: Electronic
  structure and transport properties.
\newblock {\em Phys. Rev. B}, 82:195115, 2010.

\bibitem{Karolak:PRL:2011}
M.~Karolak, D.~Jacob, and A.~I. Lichtenstein.
\newblock Orbital Kondo effect in cobalt-benzene sandwich molecules.
\newblock {\em Phys. Rev. Lett.}, 107:146604, 2011.

\bibitem{Jacob:PRB:2013}
D.~Jacob, M.~Soriano, and J.~J. Palacios.
\newblock Kondo effect and spin quenching in high-spin molecules on metal
  substrates.
\newblock {\em Phys. Rev. B}, 88:134417, 2013.

\bibitem{Anisimov:JPCM:1997a}
V.~Anisimov, A.~Poteryaev, M.~Korotin, A.~Anokhin, and G.~Kotliar.
\newblock First-principles calculations of the electronic structure and spectra
  of strongly correlated systems: Dynamical mean-field theory.
\newblock {\em J. Phys.: Condens. Matter}, 9:7359, 1997.

\bibitem{Kotliar:RMP:2006}
G.~Kotliar, S.~Y. Savrasov, K.~Haule, V.~S. Oudovenko, O.~Parcollet, and C.~A.
  Marianetti.
\newblock Electronic structure calculations with dynamical mean-field theory: A
  spectral density functional approach.
\newblock {\em Rev. Mod. Phys.}, 78:865, 2006.

\bibitem{Karolak:JoPCM:2011}
M.~Karolak, T.~O. Wehling, F.~Lechermann, and A.~I. Lichtenstein.
\newblock General DFT++ method implemented with projector augmented waves:
  electronic structure of SrVO$_3$ and the mott transition in Ca$_{2-x}$Sr$_x$RuO4.
\newblock {\em Journal of Physics: Condensed Matter}, 23:085601, 2011.

\bibitem{Pourovskii:PRB:2007}
L.~Pourovskii, B.~Amadon, S.~Biermann, and A.~Georges.
\newblock Self-consistency over the charge density in dynamical mean-field
  theory: A linear muffin-tin implementation and some physical implications.
\newblock {\em Phys. Rev. B}, 76:235101, 2007.

\bibitem{DiasdaSilva:PRB:2009}
L.~G. G.~V. Dias~da Silva, M.~L. Tiago, S.~E. Ulloa, F.~A. Reboredo, and
  E.~Dagotto.
\newblock Many-body electronic structure and Kondo properties of
  cobalt-porphyrin molecules.
\newblock {\em Phys. Rev. B}, 80:155443, 2009.

\bibitem{Korytar:JoPCM:2011}
R. Koryt\'{a}r and N. Lorente.
\newblock Multi-orbital non-crossing approximation from maximally localized
  wannier functions: the Kondo signature of copper phthalocyanine on ag(100).
\newblock {\em Journal of Physics: Condensed Matter}, 23:355009, 2011.

\bibitem{Ishida:PRB:2012}
H. Ishida and A. Liebsch.
\newblock Coulomb blockade and Kondo effect in the electronic structure of
  hubbard molecules connected to metallic leads: A finite-temperature
  exact-diagonalization study.
\newblock {\em Phys. Rev. B}, 86:205115, 2012.

\bibitem{Valli:PRB:2012}
A.~Valli, G.~Sangiovanni, A.~Toschi, and K.~Held.
\newblock Correlation effects in transport properties of interacting
  nanostructures.
\newblock {\em Phys. Rev. B}, 86:115418, 2012.

\bibitem{Ryndyk:PRB:2013}
D.~A. Ryndyk, A.~Donarini, M.~Grifoni, and K.~Richter.
\newblock Many-body localized molecular orbital approach to molecular
  transport.
\newblock {\em Phys. Rev. B}, 88:085404, 2013.

\bibitem{Baruselli:PRB:2013}
P.~P. Baruselli, M.~Fabrizio, A.~Smogunov, R.~Requist, and E.~Tosatti.
\newblock Magnetic impurities in nanotubes: From density functional theory to
  Kondo many-body effects.
\newblock {\em Phys. Rev. B}, 88:245426, 2013.

\bibitem{Knorr:PRL:2002}
N.~Knorr, M.~A. Schneider, L.~Diekh\"oner, P.~Wahl, and K.~Kern.
\newblock Kondo effect of single Co adatoms on Cu surfaces.
\newblock {\em Phys. Rev. Lett.}, 88:096804, 2002.

\bibitem{Wahl:PRL:2005}
P.~Wahl, L.~Diekh\"oner, G.~Wittich, L.~Vitali, and M.~A. Schneider.
\newblock Kondo effect of molecular complexes at surfaces: Ligand control of
  the local spin couppling.
\newblock {\em Phys. Rev. Lett.}, 95:166601, 2005.

\bibitem{Neel:PRL:2007}
N.~N\'eel, J.~Kr\"oger, L.~Limot, K.~Palotas, W.~A. Hofer, and R.~Berndt.
\newblock Conductance and Kondo effect in a controlled single-atom contact.
\newblock {\em Phys. Rev. Lett.}, 98:016801, 2007.

\bibitem{Uchihashi:PRB:2008}
T.~Uchihashi, J.~Zhang, J.~Kr\"oger, and R.~Berndt.
\newblock Quantum modulation of the Kondo resonance of Co adatoms on
  Cu/Co/Cu(100): Low-temperature scanning tunneling spectroscopy study.
\newblock {\em Phys. Rev. B}, 78:033402, 2008.

\bibitem{Soriano::2014}
M.~Soriano, D.~Jacob, and J.~J. Palacios.
\newblock in preparation.

\bibitem{ANTG}
J.~J. Palacios, D.~Jacob, P\'erez-Jim\'enez~A. J., E.~San~Fabi\'an, E.~Louis,
  and J.~A. Verg\'es.
\newblock ANT.G: Ab initio Nano Transport with Gaussian, Condensed Matter
  Theory Group, Universidad de Alicante.
\newblock http://alacant.dfa.ua.es.

\bibitem{G09}
M.~J. Frisch, G.~W. Trucks, H.~B. Schlegel, G.~E. Scuseria, M.~A. Robb, J.~R.
  Cheeseman, G.~Scalmani, V.~Barone, B.~Mennucci, G.~A. Petersson,
  H.~Nakatsuji, M.~Caricato, X.~Li, H.~P. Hratchian, A.~F. Izmaylov, J.~Bloino,
  G.~Zheng, J.~L. Sonnenberg, M.~Hada, M.~Ehara, K.~Toyota, R.~Fukuda,
  J.~Hasegawa, M.~Ishida, T.~Nakajima, Y.~Honda, O.~Kitao, H.~Nakai, T.~Vreven,
  J.~A. Montgomery, {Jr.}, J.~E. Peralta, F.~Ogliaro, M.~Bearpark, J.~J. Heyd,
  E.~Brothers, K.~N. Kudin, V.~N. Staroverov, R.~Kobayashi, J.~Normand,
  K.~Raghavachari, A.~Rendell, J.~C. Burant, S.~S. Iyengar, J.~Tomasi,
  M.~Cossi, N.~Rega, J.~M. Millam, M.~Klene, J.~E. Knox, J.~B. Cross,
  V.~Bakken, C.~Adamo, J.~Jaramillo, R.~Gomperts, R.~E. Stratmann, O.~Yazyev,
  A.~J. Austin, R.~Cammi, C.~Pomelli, J.~W. Ochterski, R.~L. Martin,
  K.~Morokuma, V.~G. Zakrzewski, G.~A. Voth, P.~Salvador, J.~J. Dannenberg,
  S.~Dapprich, A.~D. Daniels, Ö. Farkas, J.~B. Foresman, J.~V. Ortiz,
  J.~Cioslowski, and D.~J. Fox.
\newblock Gaussian∼09 revision d.01.
\newblock Gaussian Inc. Wallingford CT 2009.

\bibitem{SIESTA}
J.~M. Soler, E.~Artacho, J.~D. Gale, A.~Garc\'i­a, J.~Junquera, P.~Ordej\'on,
  and D.~Sanchez-Portal.
\newblock The SIESTA method for ab initio order-N materials simulation.
\newblock {\em Journal of Physics: Condensed Matter}, 14:2745, 2002.

\bibitem{Soriano:PRB:2014}
M.~Soriano and J.~J. Palacios.
\newblock Theory of projections with nonorthogonal basis sets: Partitioning
  techniques and effective Hamiltonians.
\newblock {\em Phys. Rev. B}, 90:075128, 2014.

\bibitem{ORegan:prb:2011}
D.~D. O'Regan, M.~C. Payne, and A.~A. Mostofi.
\newblock Subspace representations in ab initio methods for strongly correlated
  systems.
\newblock {\em Phys. Rev. B}, 83:245124, 2011.

\bibitem{Economou::1970}
E.~N. Economou.
\newblock {\em Green's functions in Quantum Physics}.
\newblock Springer Series in Solid State Physics. Springer,
  Berlin-Heidelberg-New York-Tokyo, 1970.

\bibitem{Anderson:PR:1961}
P.~W. Anderson.
\newblock Localized magnetic states in metals.
\newblock {\em Phys. Rev.}, 124:41, 1961.

\bibitem{Mahan::2000}
G.~D. Mahan.
\newblock {\em Many-Particle Physics}.
\newblock Plenum Press, New York, 3 edition, 2000.

\bibitem{Thygesen:PRB:2006}
K.~S. Thygesen.
\newblock Electron transport through an interacting region: The case of a
  nonorthogonal basis set.
\newblock {\em Phys. Rev. B}, 73:035309, 2006.

\bibitem{Taylor:PRB:2001}
J.~Taylor, H.~Guo, and J.~Wang.
\newblock Ab initio modeling of open systems: Charge transfer, electron
  conduction, and molecular switching of a C$_{60}$ device.
\newblock {\em Phys. Rev. B}, 63:121104, 2001.

\bibitem{Baer:JCP:2004}
Roi Baer, Tamar Seideman, Shahal Ilani, and Daniel Neuhauser.
\newblock Ab initio study of the alternating current impedance of a molecular
  junction.
\newblock {\em The Journal of Chemical Physics}, 120:3387, 2004.

\bibitem{Palacios::2005}
J.~J. Palacios, A.~J. P\'erez-Jim\'enez, E.~Louis, E.~SanFabi\'an, J.~A.
  Verg\'es, and Y.~Garc\'{\i}a.
\newblock Molecular electronics with Gaussian98/03.
\newblock In Jerzy Leszczynski, editor, {\em Computational Chemistry: Reviews
  of Current Trends}, volume~9. World Scientific, Singapore-New
  Jersey-London-Hong Kong, 2005.
\newblock in press.

\bibitem{Papaconstantopoulos::1986}
A.~Papaconstantopoulos.
\newblock {\em Handbook of the Band Structure of Elemental Solids}.
\newblock Plenum Press, 1986.

\bibitem{Jacob:JCP:2011}
D.~Jacob and J.~J. Palacios.
\newblock Critical comparison of electrode models in density functional theory
  based quantum transport calculations.
\newblock {\em Journal of Chemical Physics}, 134:044118, 2011.

\bibitem{Karolak:JESRP:2010}
M.~Karolak, G.~Ulm, T.~O. Wehling, V.~Mazurenko, A.~Poteryaev, and
  A.~Lichtenstein.
\newblock Double counting in LDA+DMFT - the example of NiO.
\newblock {\em J. Electron Spectrosc. Relat. Phenom.}, 181:11, 2010.

\bibitem{Czyzyk:PRB:1994}
M.~T. Czy\.{z}yk and G.~A. Sawatzky.
\newblock Local density functional and on-site correlations: The electronic
  structure of La$_{2}$CuO$_{4}$ and LaCuO$_{3}$.
\newblock {\em Phys. Rev. B}, 49:14211, 1994.

\bibitem{Aryasetiawan:PRB:2006}
F.~Aryasetiawan, K.~Karlsson, O.~Jepsen, and U.~Schonberger.
\newblock Calculations of Hubbard U from first-principles.
\newblock {\em Phys. Rev. B}, 74:125106, 2006.

\bibitem{Aryasetiawan:PRB:2004}
F.~Aryasetiawan, M.~Imada, A.~Georges, G.~Kotliar, S.~Biermann, and A.~I.
  Lichtenstein.
\newblock Frequency-dependent local interactions and low-energy effective
  models from electronic structure calculations.
\newblock {\em Phys. Rev. B}, 70:195104, 2004.

\bibitem{Haule:PRB:2010}
A.~Kutepov, K.~Haule, S.~Y. Savrasov, and G.~Kotliar.
\newblock Self-consistent $GW$ determination of the interaction strength:
  Application to the iron arsenide superconductors.
\newblock {\em Phys. Rev. B}, 82:045105, 2010.

\bibitem{Vaugier:PRB:2012}
L.~Vaugier, H.~Jiang, and S.~Biermann.
\newblock Hubbard $U$ and hund exchange $J$ in transition metal oxides:
  Screening versus localization trends from constrained random phase
  approximation.
\newblock {\em Phys. Rev. B}, 86:165105, 2012.

\bibitem{Aryasetiawan:RPP:1998}
F.~Aryasetiawan and O.~Gunnarsson.
\newblock The gw method.
\newblock {\em Rep. Prog. Phys.}, 61:237, 1998.

\bibitem{Miyake:PRB:2009}
T.~Miyake, F.~Aryasetiawan, and M.~Imada.
\newblock \textit{Ab initio} procedure for constructing effective models of
  correlated materials with entangled band structure.
\newblock {\em Phys. Rev. B}, 80:155134, 2009.

\bibitem{Haule:PRB:2001}
K.~Haule, S.~Kirchner, J.~Kroha, and P.~W\"olfle.
\newblock Anderson impurity model at finite coulomb interaction U: Generalized
  noncrossing approximation.
\newblock {\em Phys. Rev. B}, 64:155111, 2001.

\bibitem{Grewe:PRB:1981}
N.~Grewe and H.~Keiter.
\newblock Diagrammatic approach to the intermediate-valence compounds.
\newblock {\em Phys. Rev. B}, 24:4420, 1981.

\bibitem{Kuramoto:ZPB:1983}
Y.~Kuramoto.
\newblock Self-consistent perturbation theory for dynamics of valence
  fluctuations.
\newblock {\em Z. Phys. B}, 53:37, 1983.

\bibitem{Coleman:PRB:1984}
P.~Coleman.
\newblock New approach to the mixed-valence problem.
\newblock {\em Phys. Rev. B}, 29:3035, 1984.

\bibitem{Gull:RMP:2011}
E.~Gull, A.~J. Millis, A.~I. Lichtenstein, A.~N. Rubtsov, M.~Troyer, and
  P.~Werner.
\newblock Continuous-time Monte Carlo methods for quantum impurity models.
\newblock {\em Rev. Mod. Phys.}, 83:349, 2011.

\bibitem{Bulla:RMP:2008}
R.~Bulla, T.~A. Costi, and Th. Pruschke.
\newblock Numerical renormalization group method for quantum impurity systems.
\newblock {\em Rev. Mod. Phys.}, 80:3950, 2008.

\bibitem{Ruegg:PRB:2013}
A.~R\"uegg, E.~Gull, G.~A. Fiete, and A.~J. Millis.
\newblock Sum rule violation in self-consistent hybridization expansions.
\newblock {\em Phys. Rev. B}, 87:075124, 2013.

\bibitem{Meir:PRL:1992}
Y.~Meir and N.~S. Wingreen.
\newblock Landauer formula for the current through an interacting electron region 
\newblock {\em Phys. Rev. Lett.}, 68:2512, 1992.

\bibitem{Meir:PRL:1993}
Y.~Meir, N.~S. Wingreen, and P.~A. Lee.
\newblock Low-temperature transport through a quantum dot: The Anderson model
  out of equilibrium.
\newblock {\em Phys. Rev. Lett.}, 70:2601, 1993.

\bibitem{Wingreen:PRB:1994}
N.~S. Wingreen and Y.~Meir.
\newblock Anderson model out of equilibrium: Noncrossing-approximation approach
  to transport through a quantum dot.
\newblock {\em Phys. Rev. B}, 49:11040, 1994.

\bibitem{Cohen:PRL:2014}
G.~Cohen, E.~Gull, D.~R. Reichman, and A.~J. Millis.
\newblock Green's functions from real-time bold-line Monte Carlo calculations:
  Spectral properties of the nonequilibrium anderson impurity model.
\newblock {\em Phys. Rev. Lett.}, 112:146802, 2014.

\bibitem{Wahl:PRL:2004}
P.~Wahl, L.~Diekh\"oner, M.~A. Schneider, L.~Vitali, G.~Wittich, and K.~Kern.
\newblock Kondo temperature of magnetic impurities at surfaces.
\newblock {\em Phys. Rev. Lett.}, 93:176603, 2004.

\bibitem{Neel:PRL:2008}
N.~Ne\'el, J.~Kr\"oger, R.~Berndt, T.~Wehling, A.~Lichtenstein, and M.~I.
  Katsnelson.
\newblock Controlling the Kondo effect in CoCu$_n$ clusters atom by atom.
\newblock {\em Phys. Rev. Lett.}, 101:266803, 2008.

\bibitem{Vitali:PRL:2008}
L.~Vitali, R.~Ohmann, S.~Stepanow, P.~Gambardella, K.~Tao, R.~Huang,
  V.~Stepanyuk, P.~Bruno, and K.~Kern.
\newblock Kondo effect in single atom contacts: The importance of the atomic
  geometry.
\newblock {\em Phys. Rev. Lett.}, 101:216802, 2008.

\bibitem{Hay:JCP:1985}
P.~J. Hay and W.~R. Wadt.
\newblock Ab initio effective core potentials for molecular calculations -
  potentials for K to Au including the outermost core orbitals.
\newblock {\em J. Chem. Phys.}, 82:299, 1985.

\bibitem{DaSilva:PRB:2004}
J.~Da~Silva, K.~Schroeder, and S.~Bl\"ugel.
\newblock First-principles investigation of the multilayer relaxation of
  stepped Cu surfaces.
\newblock {\em Phys. Rev. B}, 69:245411, 2004.

\bibitem{Pick:PRB:2003}
\ifmmode \check{S}\else~\v{S}\fi{}. Pick, V.~Stepanyuk, A.~Baranov, W.~Hergert,
  and P.~Bruno.
\newblock Effect of atomic relaxations on magnetic properties of adatoms and
  small clusters.
\newblock {\em Phys. Rev. B}, 68:104410, 2003.

\bibitem{Nozieres:JPhys:1980}
Ph. Nozi{\'e}res and A.~Blandin.
\newblock Kondo effect in real metals.
\newblock {\em J. Physique}, 41:193, 1980.

\bibitem{Coleman:PRB:2003}
P.~Coleman and C.~P\'epin.
\newblock Singular Fermi liquid behavior in the underscreened Kondo model.
\newblock {\em Phys. Rev. B}, 68:220405, 2003.

\bibitem{Posazhennikova:PRB:2007}
A.~Posazhennikova, B.~Bayani, and P.~Coleman.
\newblock Conductance of a spin-1 quantum dot: The two-stage Kondo effect.
\newblock {\em Phys. Rev. B}, 75:245329, 2007.

\bibitem{Surer:PRB:2012}
B.~Surer, M.~Troyer, Ph. Werner, T.~O. Wehling, A.~M. L\"auchli, A.~Wilhelm,
  and A.~I. Lichtenstein.
\newblock Multiorbital Kondo physics of Co in Cu hosts.
\newblock {\em Phys. Rev. B}, 85:085114, Feb 2012.

\bibitem{Biermann:PRL:2003}
S.~Biermann, F.~Aryasetiawan, and A.~Georges.
\newblock First principles approach to the electronic structure of strongly
  correlated systems: Combining GW with DMFT.
\newblock {\em Phys. Rev. Lett.}, 90:086402, 2003.

\bibitem{Taranto:PRB:2013}
C.~Taranto, M.~Kaltak, N.~Parragh, G.~Sangiovanni, G.~Kresse, A.~Toschi, and
  K.~Held.
\newblock Comparing quasiparticle GW+DMFT and LDA+DMFT for the test bed
  material SrVO$_{3}$.
\newblock {\em Phys. Rev. B}, 88:165119, 2013.

\bibitem{Hansmann:PRB:2013}
P.~Hansmann, T.~Ayral, L.~Vaugier, P.~Werner, and S.~Biermann.
\newblock Long-range coulomb interactions in surface systems: A
  first-principles description within self-consistently combined GW and
  dynamical mean-field theory.
\newblock {\em Phys. Rev. Lett.}, 110:166401, 2013.

\end{thebibliography}

\end{document}